\documentclass[aps,pra,superscriptaddress,twocolumn,twocolumn,10pt]{revtex4-1}
\usepackage{amsmath}
\usepackage{latexsym}
\usepackage{graphicx}
\usepackage{amsfonts}
\usepackage{braket}
\usepackage{hyperref}
\usepackage{natbib}
\usepackage{amssymb}
\usepackage{nicefrac}
\usepackage{fancyhdr}
\usepackage[utf8]{inputenc}
\usepackage{dsfont}
\usepackage{colortbl}
\usepackage{xcolor}
\usepackage{subfigure}
\usepackage{psfrag}
\usepackage{tikz}
\usepackage{paralist}
\usepackage{nicefrac}
\usepackage{ulem}
\usepackage{placeins}
\usepackage{mathtools}
\usepackage{bbm}

\usetikzlibrary{arrows,shapes,decorations.pathmorphing}

\makeindex

\begin{document}

\title{Nonlinear Parity Readout with a Microwave Photodetector}

\author{M. Sch\"ondorf}
%\author{L. C. G. Govia}
%\altaffiliation[Current Address: ]{Departement of Physics, McGill University, Montreal, Quebec, Canada H3A 2T8}
\affiliation{Theoretical Physics, Saarland University, 66123
Saarbr{\"u}cken, Germany}
\author{F. K. Wilhelm}
\affiliation{Theoretical Physics, Saarland University, 66123
Saarbr{\"u}cken, Germany}

\begin{abstract}
Robust high-fidelity parity measurment is an important operation in many applications of quantum computing. In this work we show how in a circuit-QED architecture, one can measure parity in a single shot at very high contrast by taking advantage of the nonlinear behavior of a strongly driven microwave cavity coupled to one or multiple qubits. We work in a nonlinear dispersive regime treated in an exact dispersive transformation. We show that appropriate tuning of experimental parameters leads to very high contrast in the cavity and therefore to a high efficiency parity readout with a microwave photon counter or another amplitude detector. These tuning conditions are based on nonlinearity and are hence more robust than previously described linear tuning schemes. In the first part of the paper we show in detail how to achieve this for two qubit parity measurements and extend this to $N$ qubits in the second part of the paper. We also study the QNDness of the protocol.
\end{abstract}

\maketitle

\section{Introduction}
\label{sec:1}

The platform of circuit quantum electrodynamics (cQED) is a promising candidate for realizing quantum computing in circuits in a scalable architecture \cite{clarke2008superconducting,blais2007quantum,chow2012universal,blais2004cavity,you2011atomic,kelly2015state,brecht2016multilayer}. In this field, superconducting circuits are used to realize qubits. The two lowest levels of the nonlinear energy spectrum play the role of the two qubit states. Waveguides and microwave cavities allow for control and coupling of superconducting qubits \cite{you2011atomic,hofheinz2009synthesizing}. Another crucial point is readout.

Currently, readout in superconducting circuits is mostly realized using homodyne field amplitude detection \cite{blais2004cavity,eichler2012characterizing,vijay2012stabilizing,majer2007coupling,hofheinz2009synthesizing}. This scheme requires additional devices such as parametric amplifiers to measure the field amplitudes \cite{vijay2012stabilizing,vijay2011observation}. While amplifiers are readily available, they require space-consuming microwave peripherals such as circulators \cite{Pozar}. In \cite{govia2014high} we presented a scheme to readout the state of a qubit by coupling it dispersively to a driven microwave cavity and measure if the cavity is bright or dark using a microwave photon counter. It is also possible to measure multi-qubit parity states with this setup \cite{govia2015scalable}. A challenge lies in the limited sensitivity of these detectors. Effects like back reflection of incoming photons and wrong rate calibration lead to photon loss in the counter, such that one needs a relatively high number of photons to actually get a count \cite{marius2016optimization}. This can be in conflict with the applicability of the dispersive approximation \cite{boissonneault2009dispersive}. 

A way out to increase contrast at limited sensitivity is to boost the signal and use the nonlinearity of the driven cavity, similar to how it has been done in the single qubit case. In \cite{boissonneault2010improved} Boissenault et al. studied a $M$-level system dispersively coupled to a microwave cavity. They showed numerically that going to higher drive strengths where $n>n_{\rm crit}$ leads to a nonlinear behavior of the system dynamics resulting in a huge enhancement of the cavity occupation that can be used to distinguish the two logical qubit states. At the same time Bishop et al. \cite{bishop2010response} studied the same system with just two energy levels included. This amounts to a binary pre-measurement of the qubit state. They used the exact dispersive transformation \cite{carbonaro1979canonical} in a semi-classical regime to describe this phenomenon mathematically. The nonlinear effects were also demonstrated in experiment for readout of a two level system \cite{bishop2010response,bertet2012circuit}.

Here we study this transition to a nonlinear response of the cavity using the exact dispersive transformation and extend it to multiple qubits coupled to the transmission line while taking into account $M$ energy levels of the system representing the qubit. We show how the exact dispersive transformation is performed for the general case of $M$ energy levels and $N$ qubits and derive an analytical expression for the steady state photon occupation of the cavity depending on the $N$-qubit state. Our results match, in the one-qubit case,  the numerical results of \cite{boissonneault2010improved}. Analogous to their results the equations lead to a strong enhancement in the cavity occupation depending on the qubit state. This state dependence can only be seen when we include higher energy levels, since they lead to asymmetric frequency shifts of the effective cavity frequency.

We furthermore use stability analysis to derive an expression for the critical drive strength at which one can observe the strong enhancement in the cavity occupation. An important observation is that besides the qubit state, the position of this transition also depends on the detuning of the drive frequency and the bare cavity frequency. We show that with this dependency one can tune the system such that it is possible to perform any arbitrary two qubit measurement in the logical basis, including parity readout. While one drive frequency is enough to perform parity measurements for two qubits, we show that one needs $\lfloor N/2\rfloor$ different drive frequencies to extend the parity readout scheme to $N$ qubits.

The advantage of the strongly driven regime is the high contrast of about $10^5$ photons between the different states, such that even a photon detector with very low efficiency can be used to perform the scheme we present in this paper. While it seems possible to use homodyne detection at first, here the problem is that an arbitrary detuning between drive and cavity frequency is not possible, because the drive is at the same time used for readout, which also causes a phase sensitivity of the readout we do not have when we use a microwave photon counter instead. 

Another point which is crucial in this strong driven regime is the back action of the high cavity occupation on the qubit state. Since we want to perform quantum non demolition measurements (QND) to use the scheme for instance for quantum error correction \cite{Nielsen,divincenzo2009fault}, the post-measurement qubit state should be the corresponding parity eigenstate. We look at the effect of decoherence and relaxation of the qubit in this regime and show that all the appearing rates of the decoherence channels in the new frame (general dispersive frame) are of the order of the incoherent rates in the lab frame. This is important, to show that incoherent processes are not orders of magnitude greater in the frame we work in. Additionally we have to study the effect of photon leakage of the cavity on the decoherence of the qubits.

This paper is organized as follows. In Sec. \ref{sec:2} we look at the two qubit case including three energy levels per qubit. We present the system of interest, perform the transformation and calculate the photon amplitude as well as the corresponding jump positions. With these results we present a two qubit parity measurement scheme. In Sec. 3 we expand the approach of Sec. \ref{sec:2} to the general case of $N$ qubits with $M$ energy levels. We give the general form of the exact dispersive transformation and solve it in the same manner as for the two qubit case. Additionally we present different possible applications for readout, like ground state testing and multi-qubit parity measurement. In Sec. \ref{sec:4} we take a look at the ONDness of the protocol. In Sec. \ref{sec:5} we present our conclusion.

\section{Two qubit case}
\label{sec:2}

\subsection{System and Hamiltonian}
\label{sec:2_1}

\begin{figure}
\includegraphics[width=0.45\textwidth]{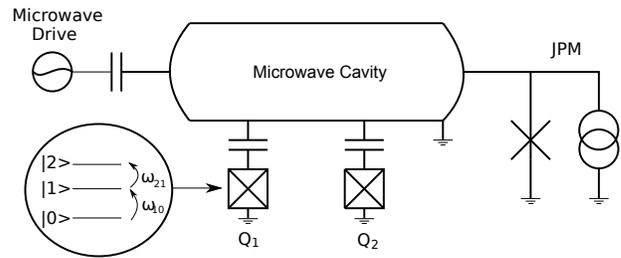}
\caption{\label{fig:JPM_TL} System schematic. $2$ Qubits are coupled to a driven microwave cavity. The existence of photons in the cavity is read out by a microwave photon counter.}
\end{figure}

Here we look at two qubits coupled to a strongly, classically driven microwave cavity. Since most of the current experiments use Transmon qubits, which have a weak anharmonicity, we will take into account 3 energy levels instead of only the two lowest qubit states. In Sec. \ref{sec:3} where we generalize the whole calculation, we expand this to the case of a general $M$ level system. The cavity is additionally coupled to a microwave photon detector, which is used to distinguish between a bright and dark cavity without detecting the phase \cite{govia2014high,govia2015scalable}. The setup is shown in Fig \ref{fig:JPM_TL}, where the photon detection is performed by the Josephson photomultiplier \cite{chen2011microwave,govia2012theory}, but in principle there are no restrictions on the type of photon detector. The bare qubit and cavity Hamiltonian $\hat H_0$ is given by
\begin{align}
\hat H_0 = \omega_c \hat a^{\dag}\hat a + \sum_{i=0}^{2} \omega_i^{(1)} \hat \Pi_i^{(1)} + \sum_{i=0}^{2} \omega_i^{(2)} \hat \Pi_i^{(2)}.
\label{2QB:Hamilton}
\end{align}
In this expression $\hat a$ and $\hat a^{\dag}$ denote the bosonic anhilation and creation operator for a cavity mode of frequency $\omega_c$, respectively and $\omega_i^{(j)}$ is the corresponding frequency of the energy level $\ket{i}^{(j)}$, where the upper index stands for the $j$-th qubit (here $j=1,2$). The operators $\hat \Pi_i^{(j)} = (\ket{i}\bra{i})^{(j)}$ are the projection operators on the $i$-th qubit energy level of the $j$-th qubit. To simplify the calculation we set $\hbar = 1$.

For later applying the exact dispersive transformation, we want to rewrite the Hamiltonian using the $\hat \sigma_z$ operators of the two dimensional subspaces
\begin{align}
\sigma_{z,i}^{(j)} = -\hat \Pi_{i-1}^{(j)} + \hat \Pi_{i}^{(j)}.
\label{sigma_z_general}
\end{align}
The result is a Hamiltonian that highlights transitions 
\begin{align}
\hat H_0 = \omega_c \hat a^{\dag} \hat a + \sum_{i=1}^{2} \tilde \omega_i^{(1)} \frac{\hat \sigma_{z,i}^{(1)}}{2} + \sum_{i=1}^{2} \tilde \omega_i^{(2)} \frac{\hat \sigma_{z,i}^{(2)}}{2},
\end{align}
where $\tilde \omega_{i}^{(j)}$ are the transformed frequencies
\begin{align}
\tilde \omega_1^{(j)} &= \frac{4\omega_{10}^{(j)}+2\omega_{21}^{(j)}}{3} \\
\tilde \omega_2^{(j)} &= \frac{2\omega_{10}^{(j)}+4\omega_{21}^{(j)}}{3},
\end{align}
with $\omega_{ij}^{(k)} = \omega_i^{(k)} - \omega_j^{(k)}$.
How to perform this transformation in general is shown in App. \ref{app:1}.
Note that for two different qubits it is $\tilde \omega_i^{(1)} \neq \tilde \omega_i^{(2)}$.

The interaction between the cavity and the two qubits in the RWA is given by a Jaynes-Cummings term \cite{jaynes1963comparison} for every allowed transition
\begin{align}
\hat H_{\rm int} &= \hat H_{\rm int, QB1} + \hat H_{\rm int, QB2} \\
&= \sum_{i=1}^{2} g_i^{(1)} \hat I_{+,i}^{(1)}  + \sum_{i=1}^{2} g_i^{(2)} \hat I_{+,i}^{(2)},
\end{align} 
with coupling strength $g_i^{(j)}$ of the particular transition, interaction operator
\begin{align}
\hat I_{\pm,i}^{(j)} = \hat a^{\dag} \hat \sigma_i^{(j)} \pm \hat a \sigma_i^{\dag(j)}
\label{Def_I}
\end{align}
and the creation and annihilation operator of the respective qubit transition
\begin{align}
\hat \sigma_{i}^{(j)} &= (\ket{i-1}\bra{i})^{(j)} \\
\hat \sigma_{i}^{\dag(j)} &= (\ket{i}\bra{i-1})^{(j)}.
\end{align}
In the case of a Transmon qubit, the coupling matrix elements $g_i^{(j)}$ between the corresponding energy levels only depend on that of the $0\leftrightarrow 1$ transition \cite{koch2007charge}
\begin{align}
g_i^{(j)} = \sqrt{i}g_1^{(j)}.
\label{coupling}
\end{align}

To get a full description of the system, we also have to take into account the classical cavity drive which is represented by the Hamiltonian \cite{WallsMilburn}
\begin{align}
\hat H_d = \epsilon \left(\hat a {\rm e}^{i\omega_d t} + \hat a^{\dag} {\rm e}^{-i\omega_d t }\right),
\label{drive}
\end{align}
with drive strength $\epsilon$ and drive frequency $\omega_d$.

Combining all terms we end up with the full Hamiltonian
\begin{align}
\hat H = \underbrace{\hat H_0 + \hat H_{\rm int}}_{\equiv \hat H_{\rm sys}} + \hat H_d.
\label{2QB:Hamilton_independent}
\end{align}

To use the setup shown in Fig. \ref{fig:JPM_TL} for readout we work in the strong dispersive regime. The dispersive regime allows to reach a QND measurement by avoiding Rabi oscillations; the strong-dispersive regime allows to resolve all spectral lines. Additionally we assume the bad cavity regime, such that we get a hierarchy of system parameter constraints, which can be satisfied in most experiments
\begin{align}
\gamma_1,\gamma_{\Phi} \ll \kappa \ll \frac{(g_i^{(j)})^2}{\omega_{i,i-1}^{(j)}-\omega_c} \ll g_i^{(j)} \ll \omega_c,
\label{condition_rates}
\end{align}
where $\kappa$ denotes the cavity decay rate and $\gamma_1$, $\gamma_{\Phi}$ the qubit decay and dephasing rate, respectively. As usual, these incoherent rates need to be smaller than those induced by the measurement in order to faithfully detect the qubit, else it would decay before the qubit is detected.

\subsection{Exact dispersive transformation}

\label{sec:2_2}

In the low photon number regime $n<n_{\rm crit}$ with $n_{\rm crit} = (\omega_{10}-\omega_c)^2/4g_1^2$, one can use the linear dispersive approximation to diagonalize $\eqref{2QB:Hamilton_independent}$. However, we want to go to regimes where $n \gg n_{\rm crit}$ and a perturbative approximation in $n/n_{\rm crit}$ fails to converge. Therefore we use a different approach, the exact dispersive transformation, which was introduced in \cite{carbonaro1979canonical} and also has been applied in circuit QED \cite{boissonneault2009dispersive,koch2007charge,bishop2010response}.

The exact dispersive transformation for two qubits has the parametric form
\begin{align}
\hat D = \exp \left(-\sum_{i=1}^2\Lambda_i^{(1)}(\hat N_i^{(1)})\hat I_{-,i}^{(1)} - \sum_{i=1}^{2} \Lambda_{i}^{(2)}(\hat N_i^{(2)})\hat I_{-,i}^{(2)}\right),
\end{align}
where the $\Lambda_i(\hat N_i^{(j)})$'s are functions of $\hat N_{i}^{(j)} = \hat a^{\dag} \hat a + \hat \Pi_i^{(j)}$. This operator denotes the excitation number of the cavity plus the $i$-th energy level. Since we are in the strong dispersive regime the $\hat N_i^{(j)}$s are approximately good quantum numbers and therefore $\Lambda$ can be seen as a scalar when performing the transformation.

Before we apply the transformation we calculate some important commutators. It is easy to show that
\begin{align}
\left[\hat I_{-,1}^{(j)}, \hat H_0 \right] &= \underbrace{\left(\omega_{10}^{(j)}-\omega_c\right)}_{\Delta_1^{(j)}} \hat I_{+,1}^{(j)} \\
\left[\hat I_{-,2}^{(j)}, \hat H_0 \right] &= \underbrace{\left(\omega_{20}^{(j)}-\omega_c\right)}_{\Delta_2^{(j)}}\hat I_{+,2}^{(j)}.
\end{align}
To simplify the notation we introduce the following nested commutator \cite{vilenkin2013representation}
\begin{align}
\textbf{ad}_{A}(B) \equiv [A,B] \hspace{0.5cm} \textbf{ad}_A^n(B) \equiv \textbf{ad}_A(\textbf{ad}_A^{n-1}(B))
\end{align}
With these commutators, we can apply the transformation on $\hat H_{\rm sys}$, using Baker Campbell Hausdorffs formula
\begin{align}
\begin{split}
\hat H_{\rm sys}^D &= \hat D^{\dag} \hat H_{\rm sys} \hat D \\
&= \hat H_0 + \sum_{j=1}^2\sum_{k=0}^{\infty} \frac{(k+1) g + \Delta_1^{(j)}\Lambda_1^{(j)}}{(k+1)!} \textbf{ad}_{\Lambda_i \hat I_{-,1}^{(j)}}^k\left(\hat I_{+,1}\right)\\
&\hspace{1 cm} +\sum_{j=1}^2\sum_{k=0}^{\infty} \frac{(k+1) g + \Delta_2^{(j)}\Lambda_2^{(j)}}{(k+1)!} \textbf{ad}_{\Lambda_i \hat I_{-,2}^{(j)}}^k \left(\hat I_{+,2}\right).
\label{2QB_Hamiton_trafo}
\end{split}
\end{align}
To get expression $\eqref{2QB_Hamiton_trafo}$ we used some properties and relations of the appearing nested commutators that we prove in Appendix \ref{app:5}. A more detailed version of this calculation is shown in Appendix \ref{app:4}.
Here we disregarded direct two photon transition terms (for instance terms proportional to $\hat a^2 \sigma_1^{\dag}\sigma_2^{\dag}$), since the probabilities for such transitions are much less than the one photon processes due to the weak anharmonicity of the Transmon potential (selection rules). It is possible to calculate a closed form of the appearing commutators which reads
\begin{align}
\begin{split}
\textbf{ad}_{\Lambda_i \hat I_{-,i}^{2k}} \left(\hat I_{+,i}\right) &= (-4)^k (\Lambda_i)^{2k} N_i^k \hat I_{+,i} \\
\textbf{ad}_{\Lambda_i \hat I_{-,i}^{2k+1}} \left(\hat I_{+,i}\right) &= -2(-4)^k (\Lambda_i)^{2k+1} N_i^{k+1} \hat \sigma_{z,i}
\label{2QB:commutator}
\end{split},
\end{align}
We put \eqref{2QB:commutator} into \eqref{2QB_Hamiton_trafo} and end up with the following expression for the transformed system Hamiltonian
\begin{align}
\begin{split}
\hat H_{\rm sys}^D = \hat H_0 &+ \sum_{k=1}^2\sum_{i=1}^2 \left[ f_1^{(j)}\left(\Delta_i^{(j)},g_i^{(j)},\Lambda_i^{(j)},N_i^{(j)}\right) \hat I_{+,1} \right.\\&\left.- 2N_q  f_2^{(j)}\left(\Delta_i^{(j)},g_i^{(j)},\Lambda_i^{(j)},N_i^{(j)}\right)\hat \sigma_{z,i}^{(j)}\right],
\end{split}
\end{align}
with 
\begin{align}
f_1 &\equiv \frac{\Delta_i\sin\left(2\Lambda_i\sqrt{N_i}\right)}{2\sqrt{N_i}} + g_i \cos\left(2\Lambda_i\sqrt{N_i}\right) \label{QB2_f1}\\
f_2 &\equiv \frac{g_i\sin\left(2\Lambda_i\sqrt{N_i}\right)}{2\sqrt{N_i}^{(j)}} + \frac{\Delta_i\left\{1-\cos\left(2\Lambda_i\sqrt{N_q}\right)\right\}}{4N_i}.
\end{align}
To obtain a diagonal system Hamiltonian we have to choose $\Lambda_i^{(j)}$ such that the $\hat I_{+,i}^{(j)}$ contribution is zero. Setting $\eqref{QB2_f1}$ equal to zero we find the following choice:
\begin{align}
\Lambda_i^{(j)} = -\frac{\arctan\left(\lambda_i^{(j)} \sqrt{N_i^{(j)}}\right)}{2 \sqrt{N_i^{(j)}}}.
\label{Lambda}
\end{align}
with $\lambda_i^{(j)} = g_i^{(j)}/\Delta_i^{(j)}$.
Finally we put this expression for $\Lambda_i^{(j)}$ into \eqref{2QB_Hamiton_trafo} and end up with the diagonal system Hamiltonian
\begin{align}
\hat H_{\rm sys}^D &= \hat H_0 
               - \sum_{j=1}^2\sum_{i=1}^2 \frac{\Delta_i^{(j)}}{2}\left(1-\sqrt{1+4\lambda_i^{(j)2}N_i^{(j)}}\right)\hat \sigma_{z,i}^{(j)},
\label{2QB:Hamilton_final}
\end{align}
This expression is exact up to the non parity conserving terms we ignored in $\eqref{2QB_Hamiton_trafo}$.
At this point we only moved $\hat H_{\rm sys}$ into the dispersive frame, but to describe the whole setup we additionally have to transform the drive Hamiltonian $\hat H_d$. Since we are interested in the regime $n \gg n_{\rm crit}$, the dive Hamiltonian stays in its original form by ignoring terms of the order $n^{-1/2}$ and $\lambda_i^2$ \cite{bishop2010response}
\begin{align}
\hat H_d^D \approx \hat H_d. 
\end{align}

For further calculations it is more convenient to work with a time independent Hamiltonian. Since $\hat H_d$ still includes a time dependence, we go into the frame rotating with the drive frequency $\hat U ={\rm e}^{-i\hat n \omega_d t}$. In this frame the drive is time independent $\hat H_d = \epsilon(\hat a^{\dag} + \hat a)$ and the system Hamiltonian just incorporates an additional frequency shift in the bare cavity part
\begin{align}
\hat H_0 = \delta_c \hat a^{\dag}\hat a+ \sum_{j=1}^2\sum_{i=1}^2 \tilde \omega_i^{(j)} \frac{\hat \sigma_{z,i}^{(j)}}{2}.
\end{align}
with $\delta_c = \omega_c - \omega_d$.
\subsection{Photon amplitude and instability}
\label{sec:2_3}

The interesting value which is crucial for the usage of the setup in Fig. \ref{fig:JPM_TL} for readout is the cavity occupation, which depends on the corresponding state of the qubit. Since \eqref{2QB:Hamilton_final} is diagonal, it is relatively easy to obtain the steady state solution of the photon amplitude.
As mentioned in \ref{sec:2_2}, we assume that the qubit occupation number is constant during the dynamics of the system, which is satisfied because of the diagonal structure of \eqref{2QB:Hamilton_final} and the strong detuning between cavity and qubit. Therefore the $\hat \sigma_{z,i}^{(j)}$s are constant, which simplifies the following calculation significantly.

As a starting point we use the Liouvillian equation to obtain an equation of motion for the annihilation operator of the cavity mode. Additionally we include an incoherent channel described by the Lindblad operator $\hat L_{\kappa} = \sqrt{\kappa}\hat a$ \cite{WallsMilburn}, which represents photon loss in the cavity with rate $\kappa$. The adjoint master equation \cite{breuer2002theory} leads to an equation of motion for the field operator $\hat a$ in the Heisenberg picture
\begin{align}
\dot{\hat a} = i\left[\hat H_{\rm sys}^D + \hat H_d, \hat a\right]- \frac{\kappa}{2} \hat a.
\end{align}
Putting in the expressions for $\hat H_{\rm sys}^D$ and $\hat H_d$, we get
\begin{align}
\dot{\hat a} = -i\left(\delta_c - \sum_{j=1}^2\sum_{i=1}^{2} \frac{g_i^{(j)}\lambda_i^{(j)}}{\sqrt{1 + 4 \lambda_i^{(j)2}  N_i^{(j)}}} \hat \sigma_{z,i}^{(j)} - i\frac{\kappa}{2}\right) \hat a - i\epsilon.
\label{2QB:field_eom}
\end{align} 
Conjugation of \eqref{2QB:field_eom} leads to the equation of motion for $\hat a^{\dag}$. We are interested in the cavity occupation in the post ringup state. Usually the steady state describes the state reached at $t\longrightarrow \infty$, but for t with $\gamma_1 t, \gamma_{\Phi}t$ $\gg$ 1 the qubit state would be completely destroyed. However, for $\kappa t$ $\gg$ 1 the system is in a pseudo steady state, where the behavior is well described by the steady state solutions. This is the reason why we work in the bad cavity limit, such that for this time $t_{\rm pseudo}$ we still meet the condition $\gamma_{1}t,\gamma_{\Phi}t$ $\ll$ 1.

\begin{figure}
\includegraphics[width=0.49\textwidth]{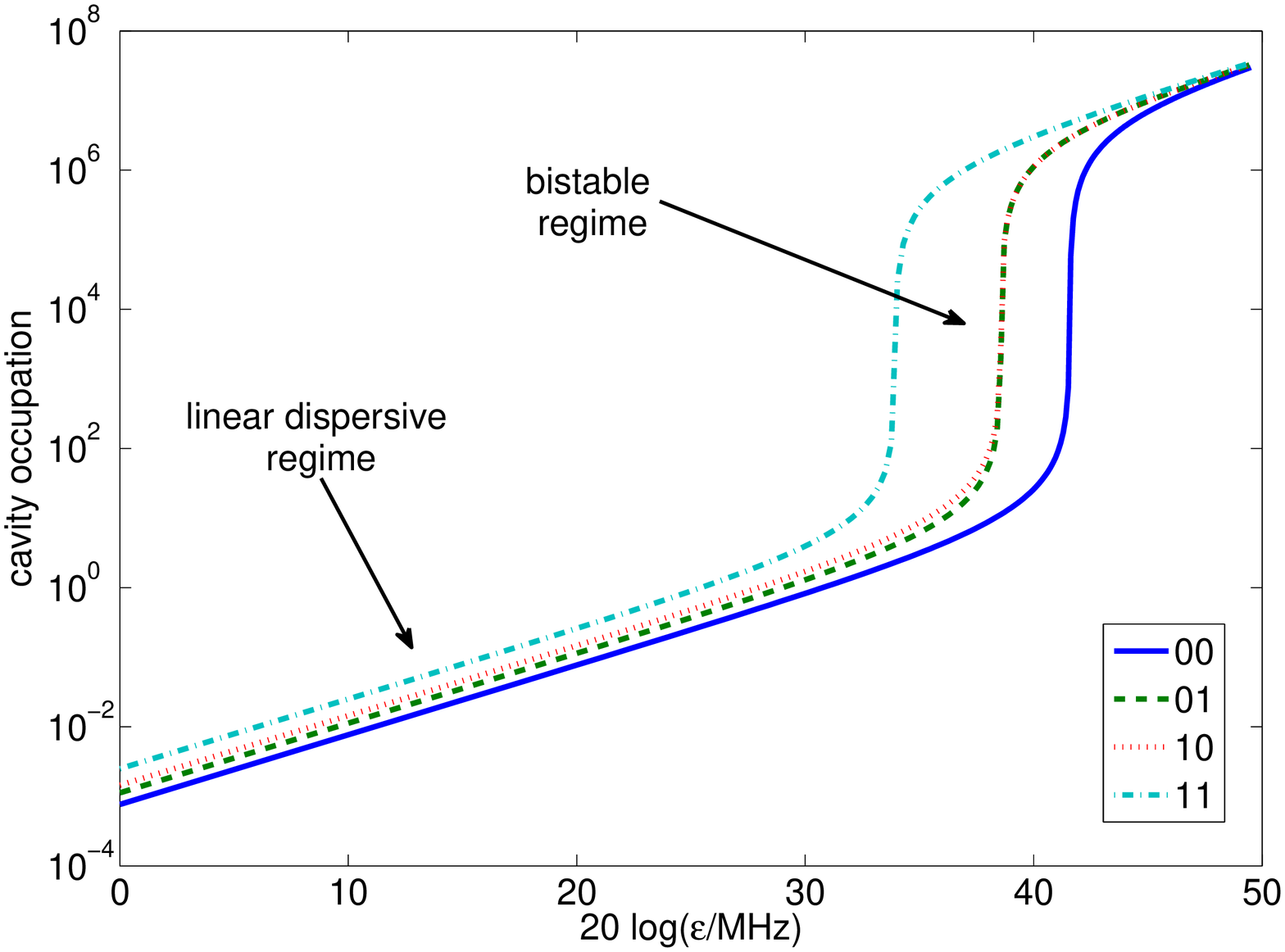}
\includegraphics[width=0.49\textwidth]{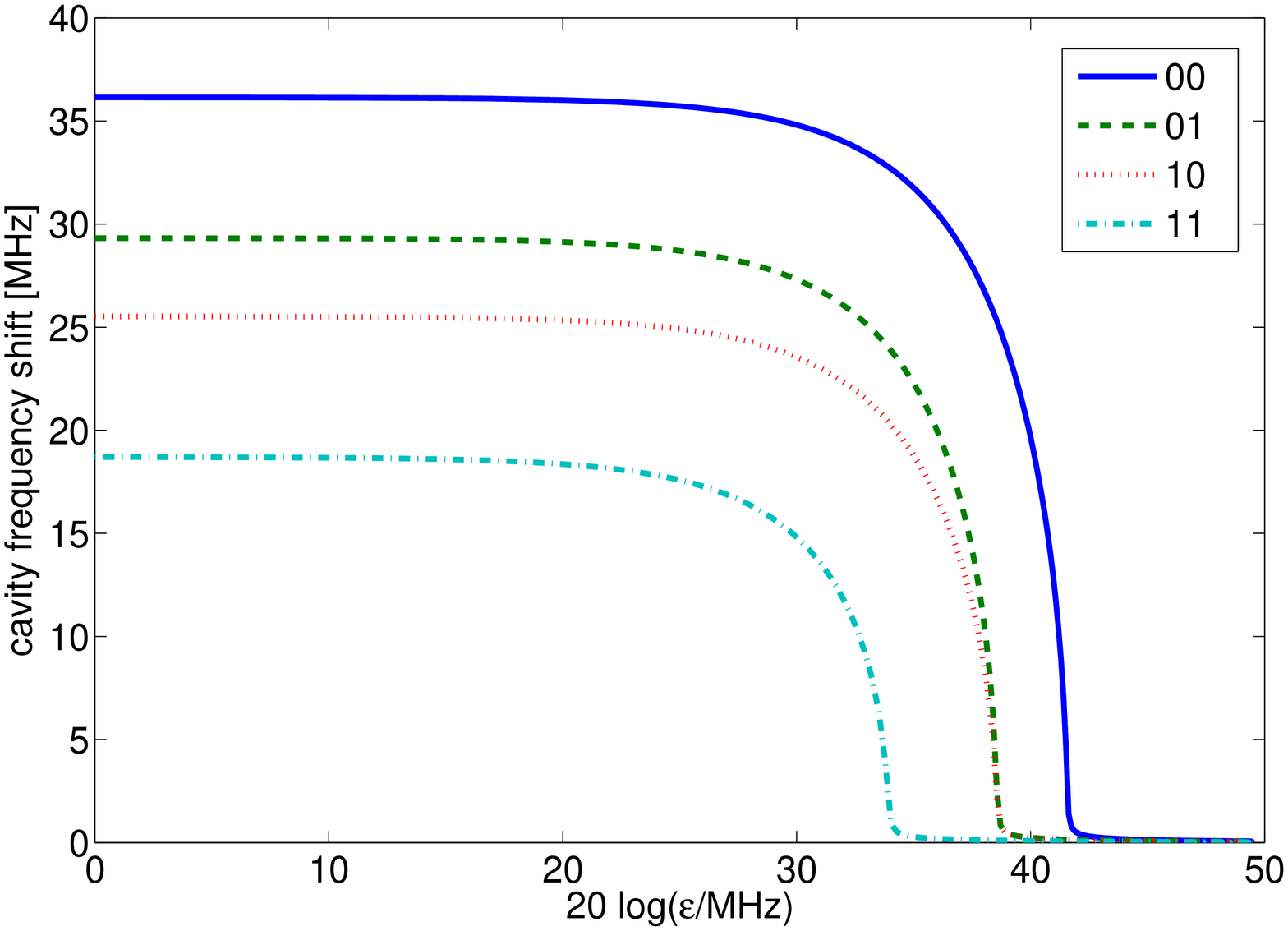}
\caption{\label{fig:2QB_1} Photon amplitude (top) and effective cavity frequency (bottom) depending on the drive strength $\epsilon$ for the four different qubit states. The parameters here are $(\omega_{10},\omega_{21},g_1)^{(1)}/2\pi = (4.297,4.071,0.12)$ GHz, $(\omega_{10},\omega_{21},g_1)^{(2)}/2\pi = (4.094,3.868,0.12)$ GHz, $\omega_c = 5.005$ GHz and $\delta_c = 0$. For every state exists a specific drive strength $\epsilon_{\rm crit}$, where the frequency rapidly jumps back to the bare cavity frequency and one observes a strong enhancement in the cavity occupation.}
\end{figure}

Setting $\dot{\hat a} = \dot{\hat a}^{\dag} = 0$ and solving both equations for $\hat a$ and $\hat a^{\dag}$ we end up with an expression for the photon occupation in the steady state
\begin{align}
n = \left<\hat a^{\dag}\hat a\right>= \frac{\epsilon^2}{\left[\delta_c - \chi\left(N_q\right)\right]^2 + \frac{\kappa^2}{4}}.
\label{2QB:photon_amplitude}
\end{align}
with nonlinear cavity frequency shift
\begin{align}
\chi\left(N_q\right) = \sum_{j=1}^2\sum_{j=1}^{2} \frac{g_i^{(j)}\lambda_i^{(j)}}{\sqrt{ 1+ 4 \lambda_i^{(j)2}N_i^{(j)}}}\sigma_{z,i}^{(j)}.
\end{align}
Note that the frequency shift itself depends on the qubit state, since it includes $\sigma_{z,i}^{(j)}$ such that the photon amplitude depends on the qubit state as well. Another crucial point is that the $N_i$s include the photon number $n$ in the cavity, such that \eqref{2QB:photon_amplitude} represents a transcendental equation. We can solve the equation iteratively and the results for some specific parameters are shown in Fig. \ref{fig:2QB_1}.

There are three regimes which can be distinguished. For low drive strengths we see a linear response of the cavity up to a critical drive strength $\epsilon_1$ and photon number $n_1$. This corresponds to the region where the system is described by the linear dispersive approximation. After that the amplitude shows a nonlinear behavior (bistable region) resulting in a strong enhancement of the photon occupation. Going to even higher drive strengths yields another critical point $(\epsilon_2,n_2)$, where the response of the cavity returns back to a linear behavior. The specific values of $\epsilon_1$ and $\epsilon_2$ depend heavily on the state of the qubit.

The effective cavity frequency on the other hand starts at a specific value which corresponds to the usual Stark shift and rapidly goes over to the bare cavity frequency in between the region $\epsilon_1<\epsilon<\epsilon_2$. In the next chapter we will see that this nonlinear behavior results from a bifurcation of the transcendental equation \eqref{2QB:photon_amplitude}. 

In the limit $n\rightarrow 0$, the expression for the frequency shift is 
\begin{align}
\lim\limits_{n \rightarrow 0}{\chi(N_q)} = \sum_{j=1}^2\sum_{i=1}^2 g_i^{(j)}\lambda_i^{(j)} \sigma_{z,i}^{(j)}.
\label{2QB:bare_chishift}
\end{align}
If we couple one qubit to the cavity and only take the two qubit states into account, we observe the linear $\chi$-shift: $\chi = \pm g_1^2/\Delta_1$ (see \cite{blais2004cavity}). Thus even though the whole calculation was performed under the assumption $n\gg n_{\rm crit}$ we still get the correct expressions for small values of $n$, such that we can assume that our equations also works well in this regime.
Another point worth to mention here is that in the case where only two levels are included the frequency shift is completely symmetric, such that the response when driving at the bare cavity frequency would be independent of the state of the qubit. This shows that the pure existence of higher levels influence the system dynamics, they do not have to be occupied at all.

\subsection{Stability anlaysis}
\label{sec:2_4}

As mentioned before, the observed strong nonlinear behavior of the photon amplitude is caused by a bifurcation of \eqref{2QB:photon_amplitude}. In between the two linear regimes (see Fig. \ref{fig:2QB_1}) equation \eqref{2QB:photon_amplitude} posses a bistable area with two attractors. Tuning the drive strength through the first bifurcation point, which appears at $\epsilon_1$, leads to a decision of the cavity dynamics between the two attractors. Which of both attractors actually describe the cavity state depends on the history of the system. In principle, small fluctuations induced by environment-assisted processes can drive transitions between the two attractors. However, as we see in Fig. \ref{fig:2QB_1} the difference in amplitude between these is about $10^6$ photons and these environmental fluctuations are assumed to be rather small. Therefore the system tends to stay in the attractor it chooses when its driven through the first bifurcation point, i.e., the transition time is exponentially long.

In this section we want to calculate the two critical points that restrict the bistable area using stability analysis similar to Drummond et al. \cite{drummond1980quantum}. It is difficult to use the full expression \eqref{2QB:photon_amplitude} for stability analysis since $n$ appears in a square root in the denominator. As we see in Fig. \ref{fig:2QB_1} the transition happens around a cavity occupation of about $10$ photons. Therefore it is a good approximation to only keep terms up to $g_i^4/\Delta_i^3$, since for $n\approx 10$ we still meet the condition $n\cdot g_i^4/\Delta_i^3 \ll 1$. Expanding the square root appearing in \eqref{2QB:Hamilton_final} up to that order we can derive an equation of motion for the field amplitudes in the same manner as in the previous section
\begin{align}
\frac{\partial}{\partial t} \begin{bmatrix}\alpha \\ \alpha^*\end{bmatrix}  = \begin{bmatrix} -i\epsilon -\left[\frac{\kappa}{2} + i h(n)\right]\alpha \\ i\epsilon - \left[\frac{\kappa}{2} - ih(n)\right]\alpha^* \end{bmatrix},
\end{align}
with $\left<\hat a\right> = \alpha$, $\left<\hat a^{\dag}\right> = \alpha^*$ and photon number $n = \alpha^*\alpha$. The function $h$ depends on the photon number $n$ and is given by
\begin{align}
h(n) = \delta_c +\sum_{j=1}^2\sum_{i=1}^2 \Delta_i^{(j)}\left(\lambda_i^{(j)2}-2\lambda_i^{(j)4}N_i^{(j)}\right) \sigma_{z,i}^{(j)}
\end{align}
With $h(n)$ the steady state condition $\dot n = 0$ can be written as
\begin{align}
\left|\epsilon\right|^2 = n\left(\frac{\kappa^2}{4} + h^2(n)\right).
\label{2QB:state_equation}
\end{align}
Now we assume small fluctuations $\Delta\alpha(t)$ around the steady state solution
\begin{align}
\alpha(t) = \alpha_0 + \Delta\alpha(t)
\end{align}
and get a linearized equation for the fluctuation
\begin{align}
\frac{\partial}{\partial t} \begin{bmatrix}\Delta\alpha \\ \Delta \alpha^*\end{bmatrix} = \textbf{A} \begin{bmatrix}\Delta\alpha \\ \Delta \alpha^*\end{bmatrix},
\end{align}
with
\begin{align}
\textbf{A} = \begin{bmatrix}i \left(n \frac{\partial h(n)}{\partial n} + h(n) \right)+ \frac{\kappa}{2}  && i \alpha_0^2 \frac{\partial h(n)}{\partial n} \\ -i \alpha_0^{*2} \frac{\partial h(n)}{\partial n} && -i\left( n \frac{\partial h(n)}{\partial n} +  h(n)\right) + \frac{\kappa}{2} \end{bmatrix}.
\end{align}
The stability of equation $\eqref{2QB:photon_amplitude}$ is then controlled by the Hurwitz criteria
\begin{align}
{\rm Tr}(\textbf{A}) &> 0 \\
{\rm Det}(\textbf{A}) &> 0. 
\label{2QB:Hurwitz}
\end{align}
If these two criteria are fulfilled, the eigenvalues of the equation are stable. Therefore the bistability can only occur if one of the two equations \eqref{2QB:Hurwitz} changes sign. Since ${\rm Tr}(A) = \kappa$ and we assume to have a cavity decay ($\kappa>0$) only the second Hurwitz criterion indicates an instability. The bistable region is restricted by the two critical points that fulfill the condition ${\rm Det}(A) = 0$, which leads to the following expression for the photon number at the critical points:
\begin{align}
n_{1/2} = \frac{-2\Delta \omega_i \mp \sqrt{\Delta\omega^2 - \frac{3}{4}\kappa^2}}{6 \chi},
\label{2QB:critical}
\end{align}
where we adopted the notation of \cite{drummond1980quantum} by defining the parameters
\begin{align}
\Delta \omega &\equiv \delta_c + \sum_{j=1}^2\sum_{i=1}^2  \Delta_i^{(j)}\left(\lambda_i^{(j)2} - 2 \lambda_i^{(j)4} N_{{\rm QB},i}^{(j)}\right)\sigma_{z,i}^{(j)} \\
\chi &\equiv -\sum_{j=1}^2\sum_{i=1}^2 2 \Delta_i^{(j)} \lambda_i^{(j)4} \sigma_{z,i}^{(j)}
\end{align}
with $N_{{\rm QB},i}^{(j)} = \left<\hat \Pi_i^{(j)}\right>$. To get the  drive strengths $\epsilon_1$ and $\epsilon_2$ corresponding to the two bifurcation points, we have to put the expression \eqref{2QB:critical} into the equation for $\epsilon$ \eqref{2QB:state_equation}
\begin{align}
\epsilon_{1/2} = \sqrt{n_{1/2}\left(\frac{\kappa^2}{4} + h^2(n)\right)}.
\label{epsilon_crit}
\end{align}
It is obvious that the photon numbers resulting from equation \eqref{2QB:critical} has to be positive, such that a bifurcation only occurs if $\Delta\omega^2 > 3\kappa^2/4$ and $ \chi\Delta\omega <0$. The first of these inequalities shows that we do not observe a nonlinear behavior if the leakage rate of the cavity is to high. On the other hand the second inequality leads to the fact that the cavity has to be in the blue detuned regime with respect to the qubit frequencies. The second condition also indicates the borders (dotted line) in Fig. \ref{fig:2QB_instability}. Note that $\Delta\omega$ as well as $\chi$ depend on the state of the two qubit subset, such that $\epsilon_1$ and $\epsilon_2$ depend on it as well, which explains the different position of the transition in Fig \ref{fig:2QB_1}.

What we know up to now is that the transition between the two attractors (which we call low and high amplitude attractor in the following) occurs at some value in the bistable area $\epsilon_{\rm crit} \in \left(\epsilon_1,\epsilon_2\right)$. The dynamics of the amplitude depend on the history of the system. Starting at a drive strength smaller than the first bifurcation point $\epsilon_1$ and slowly tune it up to higher drive strengths aims the system to stay in low amplitude attractor until it reaches the second bifurcation point $\epsilon_2$, where this attractor no longer exists and it rapidly jumps into the high amplitude attractor. On the other hand starting at higher drive strengths than the second bifurcation point $\epsilon_2$ leads to a behavior the other way round. The system stays in the high amplitude attractor until it reaches the bifurcation point $\epsilon_1$ and then rapidly "jumps" into the low amplitude attractor, since the high amplitude attractor does not exist for $\epsilon<\epsilon_1$. Therefore the dynamics of the system depend on how the tuning of the parameter $\epsilon$ is performed. 
\begin{table}
\begin{tabular}{|c||c|c|c|c|}\hline
   & 00 & 01 & 10 & 11  \\ \hline 
  $20 \log(\epsilon_{2,{\rm an}}/{\rm MHz})$ &  41.4 & 38.7 & 37.6 & 33.4  \\ \hline
  $20 \log(\epsilon_{\rm crit,plot}/{\rm MHz})$ & 41.6 & 38.6 & 37.9 & 33.2 \\ \hline
 \end{tabular}
\caption{Comparison between the analytical value of $\epsilon_2$ calculated with \eqref{epsilon_crit} and the actual vlaue of $\epsilon_{crit}$ in FIG. \ref{fig:2QB_1}. We see an almost perfect agreement.}
\label{tab:crit}
\end{table}

The values for $\epsilon_2$ for the parameters in Fig. \ref{fig:2QB_1} are given in Tab. \ref{tab:crit}. Comparing them to the actual values values of $\epsilon_{\rm crit}$ in Fig. \ref{fig:2QB_1} we see an almost perfect coincidence of $\epsilon_2$ with $\epsilon_{\rm crit}$ for all states, which is due to the fact that we started with a small photon number when we solved equation \eqref{2QB:photon_amplitude} iteratively. In a real experiment where one starts with small drive strength and slowly tunes up the drive strength the system tends to stay in the low amplitude solution for every state as long as possible, hence the transition in this case can be assumed to be very closed to $\epsilon_2$. Therefore we will assume $\epsilon_2$ to be the actual transition point of the amplitude in the following, since this is the more reasonable method in experiment.

\subsection{Application to qubit readout and 2 Qubit parity measurement}
\label{sec:2_5}

\begin{figure}
\includegraphics[width=0.45\textwidth]{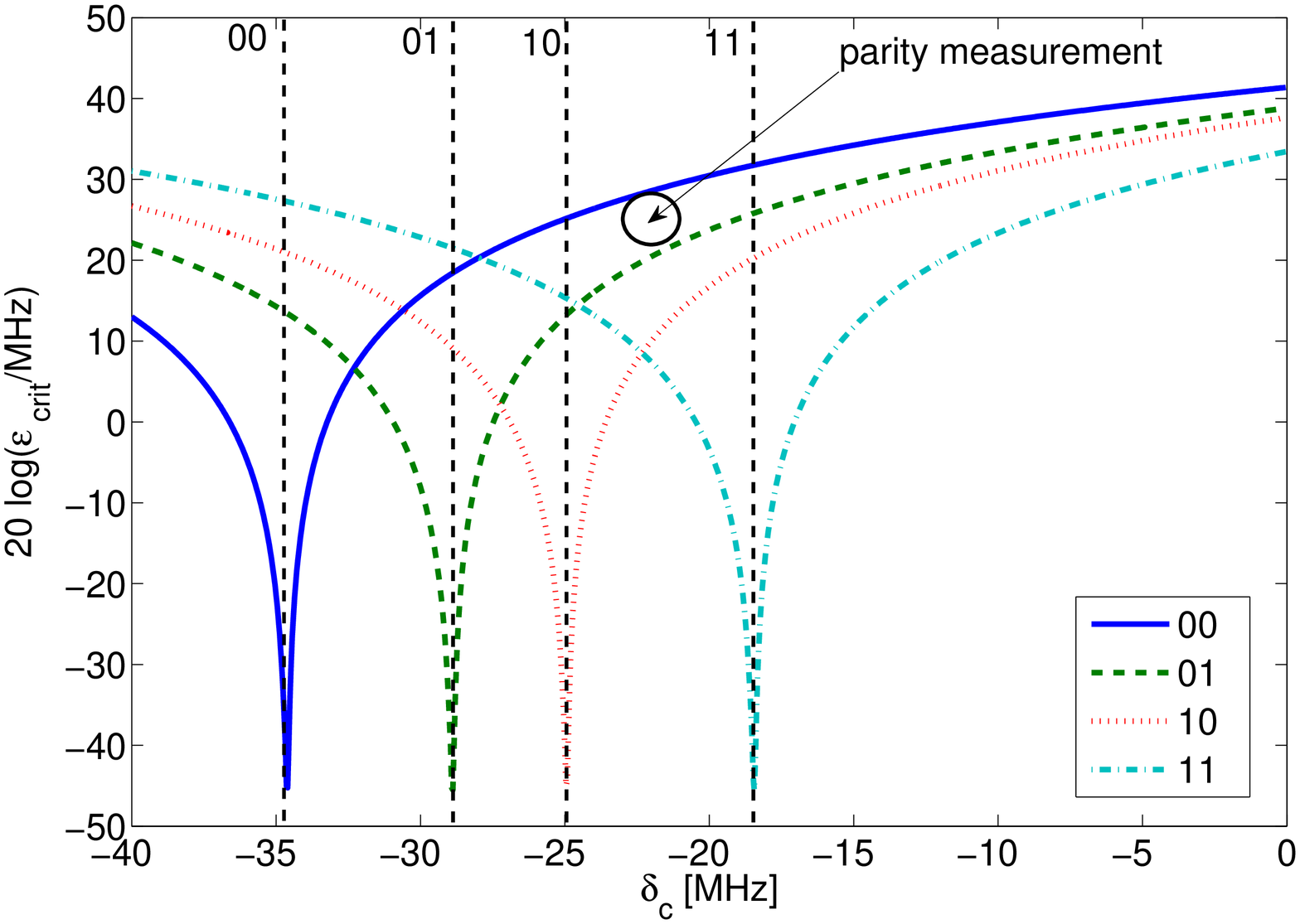}
\caption{\label{fig:2QB_instability} Critical point $\epsilon_{\rm crit} = \epsilon_2$ depending on the cavity drive detuning $\delta_c$ for the same parameters as in Fig. \ref{fig:2QB_1}. The dotted lines indicate the value of $\delta_c$, where the condition $\Delta\omega\chi<0$ is no longer fulfilled such that no bifurcation occurs if we increase $\delta_c$ to higher negative values and the system stays in the low amplitude attractor for all drive strengths. For a parity measurement we have to choose $\delta_c$ such that we are in the area where the transition to the high amplitude attractor occurs first for $\ket{01}$ and $\ket{10}$ (here $\delta_c\approx -0.02$). In this region $\ket{11}$ does not show a bifurcation and stays in the low amplitude attractor, such that for the corresponding critical drive strength $\epsilon_{\rm crit}$ the photon number for $\ket{11}$ is about $0.5$ (see inset plot). Additionally we have to drive with $\epsilon_2^{01}<\epsilon<\epsilon_2^{00}$, such that we do not reach the bifurcation point for $\ket{00}$ which assures that we also have a small photon amplitude if the system is in $\ket{00}$.}
\end{figure}

In the previous sections we assumed that we drive the cavity on resonance $\delta_c = 0$ which corresponds to homodyne detection schemes, because the drive pulse is also used for state readout \cite{mallet2009single}. Therefore it is not possible to detune drive and cavity frequency arbitrarily. Since the bifurcation points are fixed by the qubit parameters in this case, the ordering of the transition depending on the qubit states is also fixed, e.g. $\epsilon_{\rm crit}$ for the $\ket{00}$ state will always be larger than the other ones. Because of this, the usage for readout is limited. If we are interested in parity readout, there is no possibility to distinguish even from odd parity states in the case $\delta_c = 0$, since the $\epsilon_{\rm crit}$'s of the odd parity states lie in between the $\epsilon_{\rm crit}$'s of the even parity states.

Currently readout of superconducting qubits is in most cases performed by homodyne detection schemes \cite{eichler2012characterizing,vijay2012stabilizing,majer2007coupling,hofheinz2009synthesizing} where the drive and cavity frequency are in resonance or slightly detuned (heterodyne detection). Therefore the detuning between the drive and the cavity frequency is somehow fixed. However, it is also possible to use a microwave photon counter (e.g. the JPM) for readout. 
Using a microwave photon counter for the readout process of the cavity gives the possibility to arbitrarily detune the drive from the bare cavity frequency $\delta_c \neq 0$. The dependence of the critical drive strength $\epsilon_{\rm crit}$ on the detuning $\delta_c$ is shown in Fig. \ref{fig:2QB_instability}. We see that $\epsilon_{\rm crit}$ decreases if we go to higher negative values of $\delta_c$ up to a point, where the condition $\Delta\omega \chi < 0$ is no longer satisfied and we no longer have a bifurcation of \eqref{2QB:photon_amplitude}, hence the system stays in the low amplitude attractor (linear regime) over the full range of $\epsilon$ (see dotted vertical lines in Fig. \ref{fig:2QB_instability}).  

To perform two qubit parity measurements with this set up, one has to drive the system with a detuning $\delta_c$ in between the point where $\ket{11}$ goes over to a stable behavior and the point where this happens for $\ket{10}$ (circled area in Fig. \ref{fig:2QB_instability}). In this region $\epsilon_{\rm crit}$ for the odd parity states is smaller than for the $\ket{00}$ state. The $\ket{11}$ state on the other hand stays in the low amplitude solution, hence the photon number in the cavity if the system is in the $\ket{11}$ state is about $0.5$ (see Fig. \ref{fig:2QB_instability}) for the corresponding $\epsilon$. All in all with this tuning, the cavity is in a low amplitude state if the qubit is in an even parity state and vice versa.

When we take a closer look and compare the frequency shifts in Fig. \ref{fig:2QB_1} and Fig. \ref{fig:2QB_instability}, we see that the value of the detuning that gives the border between stable and unstable behavior in Fig. \ref{fig:2QB_instability} for the respective state, matches almost perfectly with the corresponding bare $\chi$-shift (at $\epsilon=0$). We will see in Sec \ref{sec:3} that this behavior can also be observed for more than two qubits. Therefore we can give an analytic expression for the optimal driving point, if one wants to perform parity measurements. The bare $\chi$-shift is given by \eqref{2QB:bare_chishift}. The optimal driving point lies in between the stability border of $\ket{11}$ and the one of $\ket{10}$ hence is given by:
\begin{align}
\omega_{d,{\rm opt}} = \omega_c + \frac{\chi_{10}+\chi_{11}}{2},
\end{align}
where $\chi_{ij}$ denotes the bare chi shift if the qubits are in the state $\ket{ij}$. The regime that can be used for parity readout hence is bounded by (area between two dotted vertical lines $11$ and $10$ in Fig. \ref{fig:2QB_instability})
\begin{align}
\omega_c + \chi_{11} < \omega_d < \omega_c + \chi_{10}.
\end{align}
Choosing a detuning in this regime, which is about $7$ MHz broad for the parameters in Fig. \ref{fig:2QB_instability}, leads to the right positions of the bifurcation points to perform two qubit parity measurements.

There is another crucial point one has to take care of when performing the measurement. We have to drive with the right frequency and the right intensity $\epsilon$ at the same time. In a real experiment, one would start with a low drive strength and tune the drive strength up into the regime $\epsilon_{\rm crit}^{01} < \epsilon < \epsilon_{\rm crit}^{00}$, hold the drive strength in this regime for $\kappa t \gg 1$ and then bring the JPM in resonance with the cavity to read out if the cavity is bright or dark, which corresponds to odd or even parity respectively.
If we tune the system in this way we have a photon occupation of about $10^5$ photons if the qubits are in an odd parity state and about $1-10$ if they are in an even parity state. Note that there can be a small difference in photon number between $\ket{11}$ and $\ket{00}$ but compared to the huge contrast between the odd and even states, this does not significantly influence the measurement. 

By tuning $\delta_c$ right one can perform any possible two qubit measurement in the logical basis, so this scheme is not restricted to parity measurement. Performing e.g. projective state measurement just needs to drive the system such that the corresponding $\epsilon_{\rm crit}$ is the lowest one. For some of the states we are in the region where the dynamics do not show a bifurcation, but this again is no problem because of the same reason as in the parity measurement scheme; the dynamics stay in the low amplitude attractor, hence the photon number is low for the corresponding $\epsilon$. 
 
Note that in the case of two identical qubits the two odd parity states would show exactly the same behavior, which means $\epsilon_{\rm crit}^{01} = \epsilon_{\rm crit}^{10}$. However, in real experiments it is often the case that the qubits have different parameters, since it is hard to produce two completely identical qubits. Therefore we assumed slightly different qubit parameters in Fig \ref{fig:2QB_instability}, and we see that if the parameters not varying to much, parity measurement can still be performed even if the qubits are not completely indistinguishable. For two qubit parity measurements we only need one drive frequency. We will see in Sec \ref{sec:3}, that this scheme can be expanded to $N$ qubit parity measurements but with a need of $\lfloor N/2\rfloor$ drive frequencies, where $\lfloor\rfloor$ denotes the floor function that maps $N$ to the next smaller integer.

\section{N Qubit case}
\label{sec:3}

\subsection{General formulation and photon amplitude}
\label{sec:3_1}

In this section we expand our result of Sec. \ref{sec:2} to $N$ qubits coupled to the readout cavity and we take into account $M$ energy levels. The bare qubit and cavity Hamiltonian $\hat H_0$ of this general case has the form
\begin{align}
\hat H_0 = \omega_c \hat a^{\dag}\hat a + \sum_{j=1}^N\sum_{i=1}^{M-1} \omega_i^{(j)} \hat \Pi_i^{(j)},
\end{align} 
where we used the same notation as in the previous section, but the upper limits of the two appearing sums are given by the number of qubits $N$ and the $M$ energy levels taken into account. Again we set $\hbar = 1$ for simplicity. For the two qubit case it was not difficult to rewrite the Hamiltonian using the $\sigma_{z,i}^{(j)}$ operators defined in Eq. \eqref{sigma_z_general}, since we just had to solve an equation system with two variables. Here we need a general transformation rule to get the corresponding Hamiltonian including only $\sigma_{z,i}^{(j)}$ operators in the bare qubit part. How to obtain this transformation is shown in Appendix \ref{app:1}. After this transformation we can write the bare Hamiltonian as
\begin{align}
\hat H_0 = \omega_c \hat a^{\dag} \hat a + \sum_{j=1}^N\sum_{i=1}^{M-1} \tilde \omega_i^{(j)} \frac{\hat \sigma_{z,i}^{(j)}}{2},
\end{align} 
with the transformation rule
\begin{align}
\tilde \omega_i^{(j)} = \sum_{k=1}^{M-1} A_{i,k}^{-1} \omega_{k,k-1},
\end{align}
where the matrix elements of $A$ are given by (see Appendix \ref{app:1})
\begin{align}
A_{i,k}^{-1} = \begin{cases}-\frac{i(k-M-2)}{M} & 1\leq i \leq k \\ -\frac{k(i-M-2)}{M} & k\leq i \leq M-1\end{cases}.
\end{align}
The interaction under the RWA leads to a Jaynes-Cummings term for every possible qubit transition summed up over all qubits
\begin{align}
\hat H_{\rm int} = \sum_{j=1}^N\sum_{i=1}^{M-1} g_i^{(j)} \hat I_{+,i}^{(j)},
\end{align}
where the definition of $\hat I_{-,i}^{(j)}$ is similar to \eqref{Def_I}. Since we are still assuming Transmon qubits, the coupling matrix elements of the respective qubit depends on the coupling rate of the corresponding $\ket{0}$ to $\ket{1}$ transition in the same manner as before (see Eq. \eqref{coupling}).

The drive Hamiltonian $\hat H_d$ does not change in the $N$ qubit case and is therefore given by \eqref{drive}. 
\begin{figure}
\includegraphics[width=0.45\textwidth]{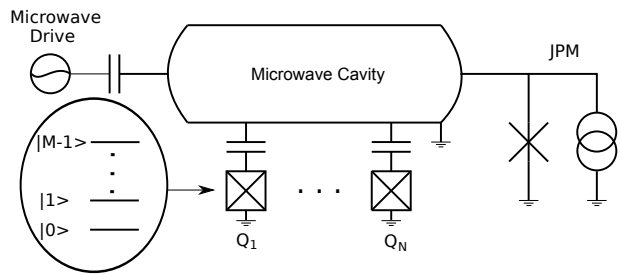}
\caption{\label{fig:N_Qubit_schematic} System schematic of the general case. $N$ Qubits are coupled to a driven microwave cavity. We take into account $M$ levels.}
\end{figure}
The exact dispersive transformation in the general case is given by 
\begin{align}
\hat D = \exp\left(\sum_{j=1}^N\sum_{i=1}^{M-1} \Lambda_i^{(j)}(\hat N_i^{(j)}) \hat I_{-,i}^{(j)}\right),
\end{align}
with $\Lambda_i^{(j)}(\hat N_i)$ defined in \eqref{Lambda}. Since $\left[\hat \sigma_{z,i},\hat \sigma_{z,i+1}\right]\neq 0$, the definition of the $\tilde \Delta_i^{(j)}$ is different for higher levels
\begin{align}
\tilde \Delta_i^{(j)} = \begin{cases} \Delta_1^{(j)} - \frac{\tilde \omega_2^{(j)}}{2}, & i = 1 \\ \Delta_{M-1}^{(j)} - \frac{\tilde\omega_{M-2}^{(j)}}{2}, & i = M-1 \\ \Delta_i^{(j)} - \left(\frac{\tilde\omega_{i-1}^{(j)}+\tilde\omega_{i+1}^{(j)}}{2}\right), & {\rm else}. \end{cases}
\end{align}

\begin{figure*}[ht]
   \centering
\subfigure[]{\includegraphics[width=0.49\textwidth]{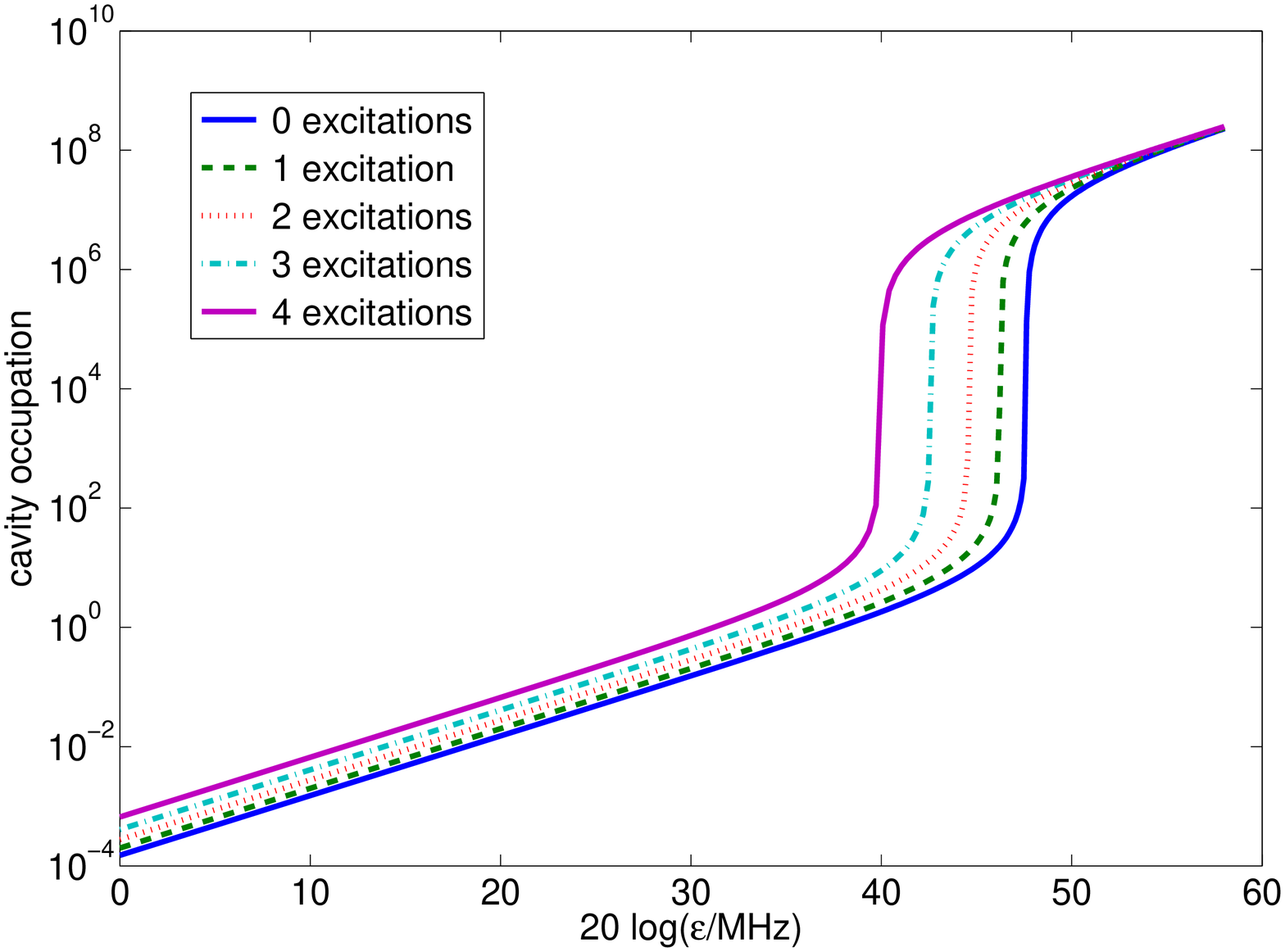}}\hfill
\subfigure[]{\includegraphics[width=0.49\textwidth]{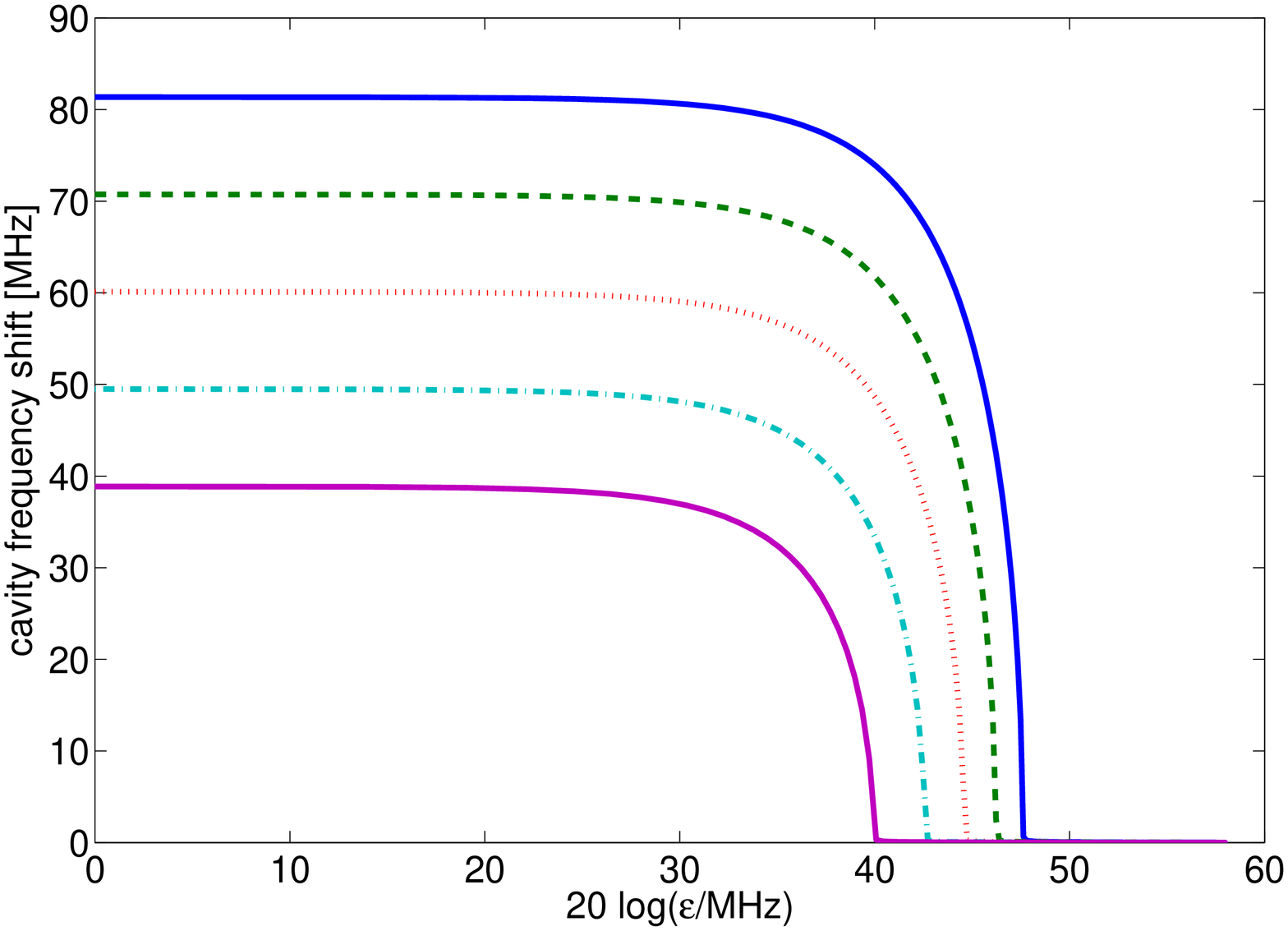}}
\caption{Photon amplitude (a) and effective cavity frequency (b) depending on the drive strength $\epsilon$. The plot is for four identical qubits including $10$ energy levels. The parameters here are $(\omega_{10},\omega_{21},g_1)^{(j)}/2\pi = (4.297,4.071,0.12)$ GHz, $\omega_c = 5.005$ GHz and $\delta_c = 0$. For every set of states with the same number of excitations there exists a specific drive strength, where the frequency rapidly jumps back to the bare cavity frequency and one observes a strong enhancement in the cavity occupation.}
\label{fig:4QB_amplitude}
\end{figure*} 
 
Applying this transformation on the system Hamiltonian is a little more difficult than in the previous section and is done in Appendix \ref{app:3}. However, the resulting Hamiltonian has a similar form, except that the appearing sums go to $N$ and $M-1$ respectively
\begin{align}
\hat H_{\rm sys}^D = \hat H_0 - \sum_{j=1}^N\sum_{i=1}^{M-1} \frac{\tilde \Delta_i^{(j)}}{2}\left(1- \sqrt{1+ 4\lambda_i^{(j)}N_i^{(j)}}\right)\hat \sigma_{z,i}^{(j)}.
\end{align}
where we again moved to the frame rotating with the drive frequency $\omega_d$, such that $\delta_c$ instead of $\omega_c$ appears in $\hat H_0$. Like before we ignored non parity conserving transitions, and the drive Hamiltonian remains in its usual form under the transformation up to order $n^{-1/2}$ and $\lambda_i^2$. 

With this Hamiltonian we can again derive equations of motion for the field amplitudes and solve the equation for the photon occupation in the cavity for the steady state resulting in 
\begin{align}
\left< n\right> = \frac{\epsilon^2}{\left[\delta_c - \chi\left(N_q\right)\right]^2 + \frac{\kappa^2}{4}}
\end{align}
which is the same expression as in section two (see Eq. \eqref{2QB:photon_amplitude}) but the appearing $\chi$ shift has more contributing terms
\begin{align}
\chi\left(N_q\right) = \sum_{j=1}^N\sum_{i=1}^{M-1} \frac{g_i^{(j)}\lambda_i^{(j)}}{\sqrt{1+4\lambda_i^{(j)2} N_i^{(j)}}}\sigma_{z,i}^{(j)}.
\label{NQB:chi_shift}
\end{align} 
The analogy of the expression we found here and the ones in the previous section indicate that it is very likely that we also observe a nonlinear behavior comparable to the two qubit case. To see that this is indeed the case, we can perform a stability analysis in the same manner as before. Doing so we get the same expression for the photon numbers at the two bifurcation points.
\begin{align}
n_{1/2} = \frac{-2\Delta \omega \mp \sqrt{\Delta\omega^2 - \frac{3}{4}\kappa^2}}{6 \chi},
\label{n_crit_N}
\end{align}
with the parameters
\begin{align}
\Delta \omega &\equiv \delta_c + \sum_{j=1}^N\sum_{i=1}^{M-1} \tilde \Delta_i^{(j)}\left(\lambda_i^{(j)2} - 2 \lambda_i^{(j)4} N_{{\rm QB},i}^{(j)}\right)\sigma_{z,i}^{(j)} \label{stablity_NQUbit1}\\
\chi &\equiv -\sum_{j=1}^N\sum_{i=1}^{M-1} 2\tilde\Delta_i^{(j)} \lambda_i^{(j)4} \sigma_{z,i}^{(j)}.
\label{stability_NQubit2}
\end{align}
The conditions that the amplitude shows a instable behavior are the same as for the two qubit case, but with the changed parameters \eqref{stablity_NQUbit1} and \eqref{stability_NQubit2}. 

Since we have more qubits and energy levels here, we get a different position for the bifurcation points for every qubit state (assuming that we have slightly different parameters for every single qubit). However, we are only interested in the occupation of the two lowest energy levels of the qubits, since they realize the two mathematical qubit states needed for quantum computation. Again it is the existence of higher levels that influence the whole system, they do not have to be occupied. Just including the two qubit levels would lead to symmetric shifts and the values for $\epsilon_{\rm crit}$ would no longer be different for all states.

The results for the four qubit case is shown in Fig. \ref{fig:4QB_amplitude}. We see that the system behaves the same as in the two qubit case. The photon amplitude shows a huge enhancement at the second bifurcation point $\epsilon_{\rm crit}$ \eqref{epsilon_crit}. Analytical expressions for the two bifurcation drive strengths can be obtained by putting the general expression for the bifurcation photon numbers $\eqref{n_crit_N}$ into the expression for the drive strength \eqref{2QB:state_equation}. In Fig. \ref{fig:4QB_amplitude} we included $10$ energy levels, which is more than the highest number of levels that are relevant in practice \cite{boissonneault2010improved}. Otherwise, we show in App. \ref{app:4} that in our case only the next lowest level which is not occupied matters, so in our case $M=3$ yields the correct results.

\subsection{Multi-qubit parity Measurements}

In Sec. \ref{sec:2_5} we have shown how to perform two qubit parity measurements in the nonlinear regime using our setup. Now we want to show that the same can be done for $N$ qubits, we just need more than one drive frequency. To perform an $N$ qubit parity measurement we need $\lfloor N/2 \rfloor$ different drive frequencies. In the $N$ qubit case the bifurcation drive strength $\epsilon_{\rm crit}$ depends on the detuning between the drive and the bare cavity frequency as well. Therefore we can again take this as an advantage to tune $\epsilon_{\rm crit}$ of the different states such that they fit for parity readout.

\begin{figure}
\includegraphics[width=.48\textwidth]{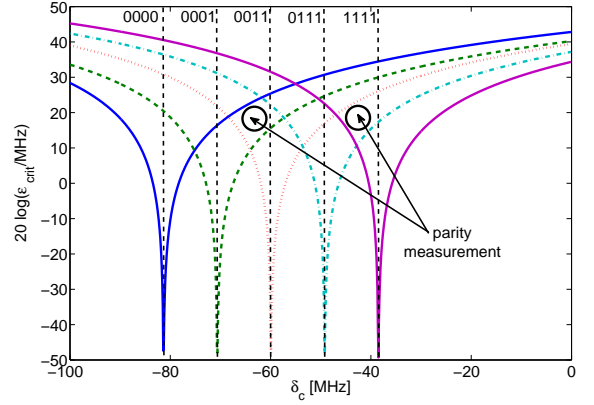}
\caption{\label{4_QB_position} Value of $\epsilon_{\rm crit} = \epsilon$ depending on the detuning between drive and cavity frequency for four identical qbuits with the same parameters as in Fig. \ref{fig:4QB_amplitude}. The dotted lines indicate the border, where the instable behavior disappears for larger detunings and the dynamics are just described by the low amplitude attractor. To measure parity we need to drive the system with two different frequencies. One such that $\delta_c$ lies in between the dotted line of $\ket{1111}$ and $\ket{0111}$ and the other one such that $\delta_c$ lies in between the dotted line corresponding to $\ket{0011}$ and $\ket{0001}$.}
\end{figure}

The values of $\epsilon_{\rm crit}$ depending on $\delta_c$ are shown in Fig. \ref{4_QB_position}, for the case of four identical qubits. Hence only qubits with different excitation number can be distinguished. We see that in the four qubit case in between the borders that restrict the instability condition lies the border of the even parity state $\ket{0011}$. Therefore it would not be possible to get $\epsilon_{\rm crit}$ for the two odd states smaller than for $\ket{0011}$ at the same time, which gives rise to the need of two different drive frequencies. One between the instability borders of $\ket{0111}$ and $\ket{1111}$, which means $\omega_c+\chi_{1111} < \omega_D^{(1)} < \omega_c + \chi_{0111}$ and the other one in between the instability borders of $\ket{0001}$ and $\ket{0011}$: $\omega_c+\chi_{0011} < \omega_D^{(2)} < \omega_c + \chi_{0001}$ as shown in Fig. \ref{4_QB_position}. By comparing the frequency shifts in Fig. \ref{fig:4QB_amplitude} and Fig. \ref{4_QB_position} we again see a coincidence with the instability borders and the bare $\chi$-shifts of the respective states which explains borders of the two drive frequencies. The optimal drive frequencies lies in the middle of the respective regime and are therefore given by
\begin{align}
\omega_{D,{\rm opt}}^{(1)} &= \omega_c - \frac{\chi_{0111}+\chi_{1111}}{2} \\
\omega_{D,{\rm opt}}^{(2)} &= \omega_c - \frac{\chi_{0001}+\chi_{0011}}{2},
\end{align}
where $\chi_{ijkl}$ denotes the bare $\chi$-shift of $\ket{ijkl}$ and can be calculated with \eqref{NQB:chi_shift}.
When we drive the system with these two frequencies, there exists a region of the drive strength 
\begin{align}
\begin{split}
{\rm max}[\epsilon_{\rm crit}^{0111}(\omega_D^{(1)}),\epsilon_{\rm crit}^{0111}(\omega_D^{(2)})] &< \epsilon \\
{\rm min}[\epsilon_{\rm crit}^{0001}(\omega_D^{(1)}),\epsilon_{\rm crit}^{0000}(\omega_D^{(2)})] &> \epsilon
\label{drive_regime}
\end{split}
\end{align}
where the dynamics of the two odd parity states are described by the high amplitude attractor and the dynamics of the even states by the low amplitude attractor (either since $\epsilon$ is smaller then the corresponding $\epsilon_{\rm crit}$, or the state no longer fulfills the condition of instability).

The calibration of the experiment can be performed in the same manner as in the one drive frequency case. First tune the two drives of the system to the right frequencies and then turn up the drive strength into the regime \eqref{drive_regime}, hold the drive strengths constant for $t \backsim 1/\kappa$ and after that bring the JPM into resonance to read out the state of the cavity.   

We can expand this measurement scheme to $N$ qubits. In this case we need $\lfloor N/2\rfloor$ different drive frequencies. Since we could show that the instability borders and the bare cavity shifts are almost identical for the two as well as for the four qubit case, we can follow that this also holds for the $N$ qubit case. The respective drive frequencies have to be in between all instability borders of odd and even states (as in the two and four qubit case). Let $\{\ket{\Psi}_i\}$ be the subset of odd parity states and $\{\ket{\Phi}_j\}$ the subset of even parity states of a $N$ qubit system, where $i$ and $j$ denote the number of excitations respectively. With this notation the optimal drive strengths are given by
\begin{align}
\omega_{D,{\rm opt}}^{(i)} = \omega_c - \frac{\chi_{\ket{\Psi}_i}+\chi_{\ket{\Phi}_{i+1}}}{2},
\end{align}
where the appearing $\chi$-shifts are again the bare $\chi$-shifts of the corresponding states. Note that in the case of an even number of qubits, it is $i=1,\ldots,N/2$ and $j=1,\ldots,N/2+1$ and in the case of an odd number of qubits $i=1,\ldots,(N+1)/2$ and $j=1,\ldots,(N+1)/2$.

\section{Is the measurement protocol QND?}
\label{sec:4}

In this section we want to take a closer look at the QNDness of our measurement protocol. One crucial point here is  to calculate the transformed incoherent rates in the new exact dispersive frame and show that they do not increase in a much faster way than the original rates, especially do not scale proportional to the photon number $n$.

Another important incoherent process is the leakage o photons out of the cavity (which is a key point for the protocol), which leads to dephasing between superpositions of equal parity states. Again it is to check if this dephasing rate is in the range of the intrinsic incoherent rates of the qubits or if it destroys the QND character of the measurement.

\subsection{Transformation of incoherent channels for $n\gg 1$}
\label{sec:4_1}

In this section we take a look at qubit intrinsic incoherent effects, such as dephasing and relaxation and how the corresponding Lindblad operators behave under the exact dispersive transformation. The consequence of such effects in the regime of small photon numbers is well studied \cite{boissonneault2009dispersive,boissonneault2008nonlinear,boissonneault2009dispersive,gambetta2008quantum,bertet2012circuit,boissonneault2012back} using the Polaron transformation, therefore we want to see how the system behaves in the regime in which we are interested, i.e. $n\gg 1$. 
Since we want to perform non-demolition measurements  our setup, it is important that the appearing dephasing and relaxation rates are not scaling with $n$ or some other parameter that is huge in our regime of interest. The optimal case is if such terms do not appear in the calculations, or at least lead to unitary perturbations of the qubits, such that we can diminish them using optimal control methods on the corresponding qubit. 

We want to focus here on the case of the first section, where we studied two three level systems coupled to a transmission line, again in the strong dispersive regime. The general case could also be calculated, but it is not necessary here since all important effects that appear in the $N$ qubit case with $M$ energy levels taken into account will also appear in this easier system. For example leakage to the fourth level will be less probable than leakage to the third one, such that we get an upper bound for all higher leakage processes. Furthermore possible interactions between the qubits induced by the high photon number in the cavity will also appear in this smaller system, if they are present, which is the reason why we include the second qubit and not just concentrate on one.

As mentioned, we are in the regime $n\gg 1$: 
\begin{align}
N_q = N_{\rm QB} + n \approx n,
\end{align}
such that $\Lambda(\hat N_q)$ acts like a scalar on qubit operators. Additionally we assume the semiclassical limit, such that
\begin{align*}
\hat a &\rightarrow \alpha \\
\hat a^{\dag} &\rightarrow \alpha^*.
\end{align*}
First we want to study relaxation of the first qubit with rate $\gamma_1$. The corresponding Lindblad operator is $\hat L_1^D = \sqrt{\gamma_1} \hat \sigma_z^D$. Therefore we have to calculate the transformation of $\hat \sigma_1$:
\begin{align}
\hat \sigma_1 &= \hat D^{\dag}\hat \sigma_1 \hat D \\
&= {\rm e}^{\Lambda_1(n)I_{1,-}+\Lambda_2(n)I_{2,-}} \hat \sigma_1 {\rm e}^{-\Lambda_1(n)I_{1,-}-\Lambda_2(N_q)I_{2,-}} \\
&= \sum_{k=0}^{\infty} \frac{1}{k!} \left(\Lambda_1^k(n)\textbf{ad}_{I_{-,1}}^k(\sigma_1) + \Lambda_2^k(n) \textbf{ad}_{I_{-,2}}^k(\sigma_1)\right).
\label{disp:expr1}
\end{align}
The expressions for the first commutator can be calculated in a closed form
\begin{align}
\textbf{ad}_{I_{-,1}}^{2k-1} \hat \sigma_1 &= 2^{2(k-1)}\sqrt{n}^{2k-1}(-1)^{k}\hat \sigma_{z,1} \label{disp:comm1}\\
\textbf{ad}_{I_{-,1}}^{2k} \hat \sigma_1 &= \begin{cases} \hat \sigma_1 & k = 0 \\ 2^{2k-1}\sqrt{n}^{2k}(-1)^{k} \hat \sigma_{x,1} & k \geq 1 \end{cases}.
\label{disp:comm2}
\end{align}
with $\sigma_{x,i} = \ket{i-1}\bra{i} + \ket{i}\bra{i-1}$.
Putting \eqref{disp:comm1} and \eqref{disp:comm2} into \eqref{disp:expr1} yields
\begin{align}
\hat \sigma_1^D = \hat \sigma_1 & - \frac{1}{2}\left( \frac{\arctan^2(2\lambda_1\sqrt{n})}{\sqrt{1+4\lambda_1^2 n}}\hat \sigma_{x,1}-\frac{2\lambda_1\sqrt{n}}{\sqrt{1+4\lambda_1^2 n}}\sigma_{z,1}\right) \\
&+ \sum_{k=0}^{\infty} \frac{1}{k!}  \Lambda_2^k(n) \textbf{ad}_{I_{-,2}}^k \hat \sigma_1
\label{disp:expr2}
\end{align}
In the same manner we get a closed expression for the second operator. First we calculate two orders
\begin{align}
\text{first order:} \hspace{1.83cm}\left[I_{2,-},\hat\sigma_1\right] &= -\alpha^* \ket{0}\bra{2} \\
\text{second order:} \hspace{0.5cm}  -\alpha^*\left[I_{2,-}, \ket{0}\bra{2}\right] &= -|\alpha|^2 \hat \sigma_2.
\end{align}
In second order there appears the commutator between $I_{-,2}$ and $\hat \sigma_2$, which can be calculated with the formula \eqref{disp:comm1} and \eqref{disp:comm2}, such that we get for $k\geq 1$
\begin{align}
\textbf{ad}_{I_{-,2}}^{2k-1} &= (-1)^{k} \sqrt{n}^{2k-1} 2^{2(k-2)} \Lambda_2^{2k-1}(n) \hat \sigma_{z,2}\label{disp:comm3} \\
\textbf{ad}_{I_{-,2}}^{2k} &= (-1)^{k} 2^{2k-3} \sqrt{n}^{2k} \Lambda_2^{2k}(n) \sigma_{x,2}\label{disp:comm4}.
\end{align}
Putting \eqref{disp:comm3} and \eqref{disp:comm4} and the expression for the first and second order into \eqref{disp:expr2}, we get the full expression for $\hat \sigma_1^D$ which yields the following expression for the corresponding Lindblad operator:
\begin{align}
\begin{split}
L_1^D &= \gamma_1 \hat \sigma_1 \\&+ \frac{\gamma_1}{2} \frac{\arctan^2(2\lambda_1\sqrt{n})}{\sqrt{1+4\lambda_1^2 n}} \hat \sigma_{x,1} + \frac{\gamma_1}{8} \frac{\arctan^4(2\lambda_2\sqrt{n})}{\sqrt{1+4\lambda_2^2 n}} \hat \sigma_{x,2} \\
&+\gamma_1\frac{\lambda_1\sqrt{n}}{\sqrt{1+4\lambda_1^2 n}} \hat \sigma_{z,1} + \frac{\gamma_1}{4} \frac{\sqrt{n}  \lambda_2 \arctan^2(2\lambda_2 n)}{\sqrt{1+4\lambda_1^2 n}}\hat \sigma_{z,2} \\
&+\frac{\gamma_1}{2} \arctan(2\lambda_2 n)\hat \sigma_1\hat\sigma_2  - \frac{\gamma_1}{8} \arctan^2(2\lambda_2 n)\hat \sigma_2
\end{split}
\end{align}
We observe as a first term the original relaxation with rate $\gamma_1$ and six additional terms. The first one is a dephasing in the $\hat \sigma_{x,1}$ basis, which leads to a mixing of the ground and first excited state. The corresponding rate has an $\arctan$ in the numerator and a term proportional to $\sqrt{n}$ in the denominator, which leads to a rate much smaller than $\gamma_1$ for $n \gg 1$ and at least smaller than $\gamma_1$ for every value of $n$. The same argument holds for the second term, which leads to a dephasing in the $\sigma_{x,2}$ basis. Here the rate is $1/4$ of the rate of the second one and therefore this term also has no crucial effect on the qubit. The third term leads to a dephasing between the ground and first excited state. The rate here is smaller than the original relaxation rate $\gamma_1$ for all values of $n$. The fourth term leads to a dephasing between the first excited state and the second excited state. Here we observe a rate which will also be in the order of the original $\gamma_1$, since we have $\sqrt{n}$ in the numerator as well as the denominator. The fifth and sixth term leads to a interactions between the second excited state and the ground and first excited state, respectively. Anyways the rates here are again at the order of the original relaxation rate $\gamma_1$. All in all we have shown that relaxation in this system leads to additional incoherent effects, but all are happening with a rate smaller or comparable with $\gamma_1$. Since we need condition \eqref{condition_rates} to perform any quantum mechanical operations with the system, nothing crucial happens here. 

Now we take a look at dephasing between the ground and first excited state. The corresponding Lindblad operator can be written as $L_{\varphi}^D = \gamma_{\varphi} \sigma_{z,1}^D$. The calculation can be performed in the same way as for $\hat \sigma_1^D$, where we need the following commutators:
\begin{align}
\textbf{ad}_{I_{-,1}}^{2k-1} \hat \sigma_{z,1} &= (-1)^{k-1} 2^{2k-1} \sqrt{n}^{2k-1} \sigma_{x,1} \\
\textbf{ad}_{I_{-,1}}^{2k} \hat \sigma_{z,2} &= (-1)^k 2^{2k} \sqrt{n}^{2k} \hat \sigma_{z,1} \\
\textbf{ad}_{I_{-,2}}^{2k-1} \hat \sigma_{z,1} &= (-1)^k 2^{2(k-1)}\sqrt{n}^{2k-1} \hat \sigma_{x,2} \\
\textbf{ad}_{I_{-,2}}^{2k} \hat \sigma_{z,1} &= (-1)^k 2^{2k-1}\sqrt{n}^{2k} \hat \sigma_{z,2},
\end{align}
which yields the following expression for the dephasing operator in the new frame:
\begin{align}
\begin{split}
L_{\varphi}^D &= \gamma_{\varphi} \hat \sigma_{z,1} \\
&- \gamma_{\varphi} \frac{\arctan^2(2\lambda_1\sqrt{n})}{\sqrt{1+4\lambda_1^2 n}} \hat \sigma_{z,1} - \frac{\gamma_{\varphi}}{2} \frac{\arctan^2(2\lambda_2\sqrt{n})}{\sqrt{1+4\lambda_2^2 n}}\hat\sigma_{z,2} \\
&- \gamma_{\varphi}\frac{2\lambda_1 \sqrt{n}}{1+4\lambda_1^2 n} \hat \sigma_{x,1} - \gamma_{\varphi}\frac{\lambda_2 \sqrt{n}}{\sqrt{1+4\lambda_2^2 n}} \hat \sigma_{x,2}.
\end{split}
\end{align}
Again we have the original dephasing term appearing in the Lindblad operator with rate $\gamma_1$. The first two additional terms lead to dephasing between the ground state and the first excited state and the first excited state and the second one, respectively. The rates are extremely small in the regime $n\gg 1$ such that we can neglect them. The last two terms lead to dephasing in the $\hat \sigma_{x,1}$ and $\hat \sigma_{x,2}$ basis, but with a rate at least smaller than the original dephasing rate $\gamma_{\varphi}$.

All in all we have shown that we do not have any relevant incoherent processes with rates higher than the relaxation and dephasing rate of the qubit, which gives the possibility to perform QND measurements in this regime. 

\subsection{Dephasing due to photon leakage}
\label{sec:4_2}
In the parity measurement protocol we assumed a cavity decay rate $\kappa$, which is greater than the intrinsic incoherent rates of the qubits. This assumption is important to reach the steady state and measure before the qubit states decay. Since states with the same parity can lead to different cavity frequency shifts (see Fig. \ref{fig:2QB_1}), the photons leaking out of the cavity carry qubit information. This leakage leads to an effective dephasing (see e.g. \cite{gambetta2006qubit}) of superpositions of parity states. We want to study this process and calculate the respective dephasing rate. Note that only even parity states cause different shifts in the cavity, since odd parity states are in the high amplitude attractor, where the frequency is exactly the bare cavity frequency for all states (see see Fig. \ref{fig:2QB_1}). Therefore we only study the dephasing between even parity states. 

In the measurement protocol, the drive strength is chosen such that the system stays in the low amplitude attractor for even parity states. Here the behavior is still linear (see \ref{fig:2QB_1}), hence we can approximate the Hamiltonian \eqref{2QB:Hamilton_final}
\begin{align}
\hat H \approx \hat H_0 + \sum_{j=1}^2\sum_{i=1}^2 \frac{\left(g_i^{(j)}\right)^2}{\Delta_i^{(j)}} \hat a^{\dag}\hat a \hat \sigma_{z,i}^{(j)} + \hat H_d,
\end{align}
where we assumed that we are already in the frame rotating with the drive frequency. The incoherent evolution of the density matrix is described by the Lindbladian master equation
\begin{align}
\begin{split}
\dot \rho &= -i \left[H,\rho\right] + \kappa \mathcal{D}[\hat a]\rho \\&\hspace{0.3cm}+ \sum_{j=1}^2\sum_{i=1}^2\gamma_{\rm 1,i} \mathcal{D}[\hat \sigma_i^{(j)}]+\sum_{j=1}^2\sum_{i=1}^2\gamma_{\rm \Phi,i} \mathcal{D}[\hat \sigma_{z,i}^{(j)}]
\end{split},
\label{deph:Lindblad}
\end{align}
which includes three incoherent processes, photon loss of the cavity with rate $\kappa$, relaxation with rate $\gamma_{1,i}$ and dephasing with rate $\gamma_{\Phi,i}$. We assume the intrinsic incoherent rates to be equal for all qubits. The qubit-cavity density matrix of an equal superposition of even parity states can be written down as (for simplicity we label $\ket {00} = \ket{0}$ and $\ket{11}=\ket{1}$)
\begin{align}
\hat \rho = \hat \rho_{00}^c\ket{0}\bra{0} + \hat \rho_{01}^c\ket{0}\bra{1} + \hat \rho_{10}^c \ket{1}\bra{0} + \hat\rho_{11}^c\ket{1}\bra{1},
\label{deph:density}
\end{align}
where $\hat \rho_{ij}^c$ describe the field part of the density matrix. Putting this density matrix expression into \eqref{deph:Lindblad} we get equations of motion for the density matrix elements $\hat \rho_{ij}$. These can be solved using the positive-P representation leading to the time evolution of the density matrix
\begin{align}
\hat \rho(t) &= \sum_{i,j=0}^1 c_{ij}(t) \ket{i}\bra{j}\otimes \ket{\alpha_i(t)}\bra{\alpha_j(t)} 
\label{deph:density_time}
\end{align}
For the detailed calculation see Appendix \ref{app:5}. The induced qubit dephasing is described by the parameter
\begin{align}
c_{10}(t) = \frac{a_{10}(t)}{\braket{\alpha_1(t)|\alpha_0(t)}},
\end{align}
with 
\begin{align}
\begin{split}
a_{10} &= a_{10}(0){\rm e}^{\left[-(\gamma_2 + i\tilde\omega)t\right]}{\rm e}^{\left[-i4(\chi_1-\chi_2/2)\int_0^t \alpha_1(t')\alpha_0^*(t'){\rm d}t'\right]}
\end{split}
\end{align}
and 
\begin{align}
\alpha_1(t) &= \alpha_1^s + {\rm e}^{\left[-(\frac{\kappa}{2}+i2(\chi_1-\chi_2)+i\delta_c)t\right]}\left(\alpha_1(0)-\alpha_1^s\right) \\
\alpha_0(t) &= \alpha_0^s + {\rm e}^{\left[-(\frac{\kappa}{2}-i2\chi_1+i\delta_c)\right]}\left(\alpha_0(0)-\alpha_0^s\right).
\end{align}
The steady state values of the field operators are given by 
\begin{align}
\alpha_1^s &= \frac{-i \epsilon}{\kappa/2+i2(\chi_1-\chi_2)+i\delta_c}\\
\alpha_0^s &= \frac{-i \epsilon}{\kappa/2-i2\chi_1+i\delta_c}.
\end{align}
The results are similar to \cite{gambetta2006qubit}, but get an additional $\chi_2$ contribution from the third energy level. This leads to a no longer symmetric dependence of the dephasing rate on the detuning $\delta_c$, which can be explained by the asymmetric frequency shift of the cavity (in the two level case the shift is $\pm \chi$, hence symmetric).
Note that we assumed identical qubits in the derivation such that the appearing linear shifts read $\chi_i = g_i^2/\Delta_i$. The measurement should be performed in a pseudo steady state, when system dynamics are almost zero. Therefore we assume the limit $\kappa t$ $\gg$ 1, where the photon leakage induced dephasing rate can be written as 
\begin{align}
\Gamma_{\Phi} = -4(\chi_1-\chi_2/2){\rm Im}\left\{\alpha_1^s\alpha_0^s*\right\}.
\end{align}
Putting all together we finally get the following expression for the dephasing rate induced by photon leakage of the cavity
\begin{align}
\Gamma_{\Phi} = \frac{4\kappa\epsilon^2\chi_2}{(\frac{\kappa^2}{4}+\delta_c^2+2\delta_c\chi_2+4\chi_1^2-4\chi_1\chi_2)^2+\kappa^2\chi_2^2}
\end{align}
As we see in Fig. \ref{fig:2QB_instability}, for the two qubit parity measurement protocol the detuning is $\delta_c \approx -20$ MHz and the corresponding drive strength is $\epsilon \approx 10$ MHz. These parameters lead to an effective dephasing of the qbuit with rate $\Gamma_{\Phi} \approx 9$ kHz. Since the cavity decay is assumed to be in the range of a few MHz, we can assume that a pseudo steady state is reached before the cavity photon loss has a significant decoherence effect on the superposition of equal parity states which leaves the measurement QND. For the $N$ qubit case the result is similar, but since there are more than one drive strengths needed, we also have more dephasing channels. However, they are all in the range of a few kHz, hence even adding all of them up does not lead to a significant dephasing.

Note that one could as well use the Polaron Transformation in the manner of \cite{boissonneault2012back} to calculate the repsective dephasing rate 
\section{Conclusion}
\label{sec:5}

In conclusion we have derived a mathematical description of $N$ superconducting qubits coupled dispersively to a microwave cavity by generalizing the exact dispersive transformation. We have obtained that our system of interest shows a nonlinear behavior for a critical drive strength, that results in a huge enhancement ($\sim 10^5$ photons) of the photon occupation in the microwave cavity. This critical drive strength depends on the qubit state and can therefore be used for high efficiency state readout. 

Furthermore we have shown this the state dependent critical drive strength can be varied by the detuning between cavity and drive. Due to this dependence it is possible to perform various high efficiency measurements including multi qubit parity readout, using a microwave photon counter to measure the cavity occupation and we have shown how to tune the system to realize these measurements. We gave expressions for the drive frequencies to perform mutli qubit parity measurements, where one needs $\lfloor \frac{N}{2} \rfloor$ different drive frequencies to measure the parity of $N$ coupled qubits.

Additionally we studied the effect of relaxation and dephasing in the high occupation regime and have shown that the appearing incoherent rates are smaller or equal to the original rates. Also the photon leakage rate of the cavity does not lead to a fast decay of qubit coherence. There are some other incoherent processes that could be considered (like broadening due to photon number variations of equal parity states), but they are all assumed to be not as significant as the studied process. This makes the presented protocol to a candidate for high contrast QND parity readout.

\section{Acknowledgments}

We thank Luke Govia, Caleb Howington, Lukas Theis and Britton Plourde for fruitful discusions.
This research is supported by U.S. Army Research Office Grant No. W911NF-15-1-0248.
%%%%%%%%%%%%%%%%%%%%%%%%%%%%%%%%%%%%%%%%%%%%%%%%%%%%%%%%%%%%%%%%%%%%%%%%%%%%%%%%%%%%%%%%%%%%%%%%%%%%%%%%

\begin{appendix}

\section{Frequency transformation in the bare Hamiltonian}
\label{app:1}
We start with the bare qubit and cavity Hamiltonian
\begin{align}
\hat H_0 = \omega_c \hat a^{\dag} \hat a + \sum_{j=1}^N\sum_{i=0}^{M-1} \omega_i^{(j)} \ket{i}\bra{i}^{(j)}.
\label{App:Hamilton1}
\end{align}
The goal is to transform the state projection operators $\ket{i}\bra{i}^{(j)}$ into the operators
\begin{align}
\hat \sigma_{z,i}^{(j)} = -\ket{i-1}\bra{i-1}^{(j)} + \ket{i}\bra{i}^{(j)}, 
\end{align}
such that we end up with a Hamiltonian of the form 
\begin{align}
\hat H_0 = \delta_c \hat a^{\dag}\hat a + \sum_{j=1}^{N}\sum_{i=1}^N \tilde \omega_i^{(j)} \frac{\hat \sigma_{z,i^{(j)}}}{2}.
\label{App:Hamilton2}
\end{align}
Comparing \eqref{App:Hamilton1} and \eqref{App:Hamilton2} we get the following transformation rule for the frequencies:
\begin{subequations}
\begin{align*}
2\omega_0^{(j)} &= -\tilde \omega_1^{(j)}  + \sum_{k=1}^{M-1} \beta_j^{(j)} \tilde \omega_j^{(j)} \\
2\omega_1^{(j)} &= \tilde \omega_1^{(j)} - \tilde \omega_2^{(j)} +  \sum_{k=1}^{M-1} \beta_j^{(j)} \tilde \omega_j^{(j)} \\
2\omega_2^{(j)} &= \tilde \omega_2^{(j)} - \tilde \omega_3^{(j)} +   \sum_{k=1}^{M-1} \beta_j^{(j)} \tilde \omega_j^{(j)} \\
&\vdots \\
2\omega_{M-2}^{(j)} &= \tilde \omega_{M-2}^{(j)} - \tilde \omega_{M-1}^{(j)} +  \sum_{k=1}^{M-1} \beta_j^{(j)} \tilde \omega_j^{(j)} \\
2\omega_{M-1}^{(j)} &= \tilde \omega_{M-1}^{(j)} +  \sum_{k=1}^{M-1} \beta_j^{(j)} \tilde \omega_j^{(j)},
\end{align*}
\end{subequations}
where $\beta_j^{(j)}$ is an arbitrary complex number. The last term in the equations comes from the fact that we can add an arbitrary vacuum contribution to the Hamiltonian in every qubit subspace without changing the system dynamics. Subtracting the second equation from the first and so on for every pair of neighboring equations leads to 
\begin{subequations}
\begin{align*}
2\omega_{10}^{(j)} = \omega_1^{(j)} - \omega_0^{(j)} &= 2 \tilde \omega_1^{(j)} - \tilde \omega_2^{(j)} \\
2\omega_{21}^{(j)} &= - \tilde \omega_1^{(j)}+ 2\tilde\omega_2^{(j)}-\tilde\omega_3^{(j)} \\
2\omega_{32}^{(j)} &= -\tilde \omega_2^{(j)} + 2\tilde \omega_3^{(j)} - \tilde\omega_4^{(j)} \\
&\vdots\\
2\omega_{M-2,M-3}^{(j)} &= -\tilde \omega_{M-3}^{(j)} + 2 \tilde\omega_{M-2}^{(j)} - \tilde\omega_{M-1}^{(j)} \\
2\omega_{M-1,M-2}^{(j)} &= -\tilde \omega_{M-2}^{(j)} + 2 \tilde \omega_{M-1}^{(j)}.
\end{align*}
\end{subequations}
This can be written down in a matrix representation
\begin{align}
2\omega_{i,i-1}^{(j)} = \sum_{k=1}^{M-1}  A_{i,k} \tilde \omega_k^{(j)},
\end{align}
with the transformation matrix
\begin{align}
A = \begin{pmatrix} 2 & -1 & 0 & 0 & 0 & 0 & \cdots & 0 \\ -1 & 2 & -1 & 0 & 0 & 0 & \cdots & 0 \\ 0 & -1 & 2 & -1 & 0 & 0 & \cdots & 0 \\  &  &  & \ddots & \ddots & \ddots & & \vdots \\ 0 & & \cdots & \cdots & &-1 & 2 & -1 \\ 0 & & \cdots & \cdots & & &-1 & 2 \end{pmatrix}.
\end{align}
This is a Töplitz matrix with $c=b=-1$ and $a=2$. The inverse of this kind of Töplitz matrix can be found in literature and is given by
\begin{align}
A^{-1} = \begin{cases} P_{ij} & 1 \leq i \leq j \\ Q_{ij} & j \leq i \leq n \end{cases},
\end{align}
with
\begin{align*}
P_{ij} &= -\frac{i(j-n-1)}{(n+1)} \\
Q_{ij} &= -\frac{j(i-n-1)}{(n+1)}.
\end{align*}
Combining all these results the frequencies $\tilde\omega_i^{(j)}$ can be calculated as
\begin{align}
\tilde \omega_i^{(j)} = \sum_{k=1}^{M-1} 2 A_{i,k}^{-1} \left(\omega_k^{(j)}-\omega_{k-1}^{(j)}\right).
\end{align}

\section{Exact dispersive transformation}
\label{app:3}

In this section of the appendix we show how to diagonalize the Hamiltonian 
\begin{align}
\hat H_{\rm sys} = H_0 + H_{\rm int}.
\end{align}
To do so we apply the unitary transformation 
\begin{align}
\hat D = \exp\left[-\sum_{j=1}^{N}\sum_{i=1}^{M-1} \Lambda_i^{(j)}(\hat N_i) \hat I_{-,i}^{(j)}\right],
\label{app:trafo}
\end{align}
where $\Lambda_i^{(j)}(N_i)$ is a scalar function of $\hat N_i$. The excitation number of every subspace ($\hat N_i$) is assumed to be constant, such that $\Lambda$ acts like a scalar on the system Hamiltonian. We use the Baker Campbell Hausdorff formula 
\begin{align}
 {\rm e}^A B {\rm e}^{-A} = \sum_{m=0}^{\infty} \frac{1}{m!}\textbf{ad}_A^n B.
\end{align}
Before we transform $\hat H_{\rm sys}$ we calculate some important commutators. It is easy to show that
\begin{align}
\left[\hat I_{-,i}^{(j)},\hat H_0\right] = \tilde\Delta_i^{(j)} I_{+,i},
\label{app:commutator}
\end{align}
where
\begin{align}
\tilde \Delta_i^{(j)} = \begin{cases} \delta_1^{(j)} - \frac{\tilde \omega_2^{(j)}}{2}, & i = 1 \\ \Delta_{M-1}^{(j)} - \frac{\tilde \omega_{M-2}^{(j)}}{2}, & i = M-1 \\ \Delta_i^{j} - \left(\frac{\tilde \omega_{i-1}^{(j)}+\tilde\omega_{i+1}^{(j)}}{2}\right), & {\rm else}. \end{cases}
\end{align}
and the definitions of Sec. \ref{sec:2_1}. With this commutator relation, we can calculate the transformation
\begin{align}
&\hat D^{\dag} H_{\rm sys} \hat D = \hat H_0 + \sum_{j=1}^N\sum_{i=1}^{M-1} g_i^{(j)} \hat I_{-,i}\\ &+ \sum_{k=1}^{\infty} \frac{1}{k!}\sum_{j=1}^N \sum_{i=1}^{M-1} \textbf{ad}_{\Lambda_i^{(j)}I_{-,i}}^{k} \left(\hat H_0 + \sum_{l=1}^{M-1} g_l^{(j)}\hat I_{-,l}\right)  \\
&= \hat H_0 + \sum_{k=0}^{\infty} g_i^{(j)}\frac{k+1}{(k+1)!} \textbf{ad}_{\Lambda_i^{(j)}I_{-,i}}^{k} \left(\hat I_{+,i}^{(j)}\right) \\ &+ \sum_{k=0}^{\infty} \sum_{j=1}^N\sum_{i=1}^{M-1} \textbf{ad}_{\Lambda_i^{(j)}I_{-,i}}^{k} \left(\hat I_{+,i}\right) \\
&= \hat H_0 + \sum_{N,M} \sum_{k=1}^{\infty} \frac{(k+1)g_i^{(j)} + \tilde\Delta_i^{(j)} \Lambda_i^{(j)}}{(k+1)!} \textbf{ad}_{\Lambda_i^{(j)}I_{-,i}} \left(\hat I_{+,i}\right).
\label{app:trafo_expression}
\end{align}
Here we made two steps, where we take use of the two relations we proof in Sec. \ref{app:4}. The firs one is, that we splitted the two parts of the Hamiltonian on the right entry of the nested commutators, so we assumed
\begin{align}
\textbf{ad}_{\sum_{i=1}^{M-1} \Lambda_i \hat I_{-,i}}^k \hat H_0 = \sum_{i=1}^{M-1} \textbf{ad}_{\Lambda_i \hat I_{-,i}} \hat H_0
\label{help1}
\end{align}
which is true due to the fact that $\left[\hat I_{-,i},\hat I_{-,j}\right] = 0$ for $i\neq j$, if we ignore non parity conserving terms and $\left[\hat I_{-,i}\left[\hat H_0 , \hat I_{-,j}\right]\right] = 0$. The last relations is true since $\left[\hat H_0,I_{-,j}\right] \propto \hat I_{-,j}$. Using this relations and the proof in Appendix. \ref{app:4} , \eqref{help1} is true.\\
On the other hand, we used that
\begin{align}
\textbf{ad}_{\sum_{i=1}^{M-1} \Lambda_i\hat I_{-,i}}^k\left(\sum_{j=1}^{M-1} \hat I_{+,j}\right) = \sum_{i=1}^{M-1} \textbf{ad}_{\Lambda_i\hat I_{-,i}}^k \hat I_{+,i},
\end{align}
which is true due to the proof in Appendix \ref{app:4} and with the relation $\left[I_{-,i},\hat I_{+,j}\right] = 0$ for $i\neq j$ again up to non parity conserving terms. 

There only appears one commutator in the expression and it can easily be calculated to be
\begin{align}
\textbf{ad}_{\Lambda_i^{(j)}I_{-,i}}^{2n}\left(\hat I_{+,i}^{(j)}\right) &= (-4)^n \Lambda_i^{(j)2n} N_i^n I_{+,i} \\
\textbf{ad}_{\Lambda_i^{(j)}I_{-,i}}^{2n+1}\left(\hat I_{+,i}^{(j)}\right) &= -2(-4)\Lambda_i^{(j)2n+1} N_i^{n+1}  \sigma_{z,i}.
\label{app:commutator2}
\end{align}
Putting \eqref{app:commutator2} into \eqref{app:trafo_expression} we end up with:
\begin{widetext}
\begin{align}
&\hat H_{\rm sys}^D = \hat H_0 \\&+ \sum_{j=1}^N\sum_{i=1}^{M-1}\left(\left[\frac{\tilde \Delta_i^{(j)}\sin\left(2\Lambda_i^{(j)}\sqrt{N_i}\right)}{2\sqrt{N_i}} + g_i^{(j)} \cos\left(2\Lambda_i^{(j)}\sqrt{N_i}\right)\right]I_{+,1} \right.\\
&\left.- 2N_i \sigma_{z,i} \left[\frac{g_i^{(j)}\sin\left(2\Lambda_i^{(j)}\sqrt{N_i}\right)}{2\sqrt{N_i}} + \frac{\tilde\Delta_i^{(j)}\left\{1-\cos\left(2\Lambda_i^{(j)}\sqrt{N_i}\right)\right\}}{4N_i}\right]\right).
\end{align}
\end{widetext}
To get a diagonal Hamiltonian we have to define the scalar functions $\Lambda_i^{(j)}$ as follows:
\begin{align}
\Lambda_i^{(j)}(N_i) = -\frac{\arctan\left(2\lambda_i^{(j)}\sqrt{N_i}\right)}{2 \sqrt{N_i}}.
\end{align}
With this choice for $\Lambda_i^{(j)}$ we end up with a diagonal system Hamiltonian
\begin{align}
\hat H_{\rm sys}^D =  \delta_c \hat a^{\dag} \hat a  + \hat H_0 - \sum_{j=1}^N\sum_{i=1}^{M-1} \frac{\tilde \Delta_i^{(j)}}{2}\left(1- \sqrt{1+ \frac{g_i^{(j)}}{\tilde \Delta_i^{(j)}}N_i}\right)
\end{align}

\section{The effect of higher energy levels}
\label{app:4}

\begin{figure}
\includegraphics[width=0.45\textwidth]{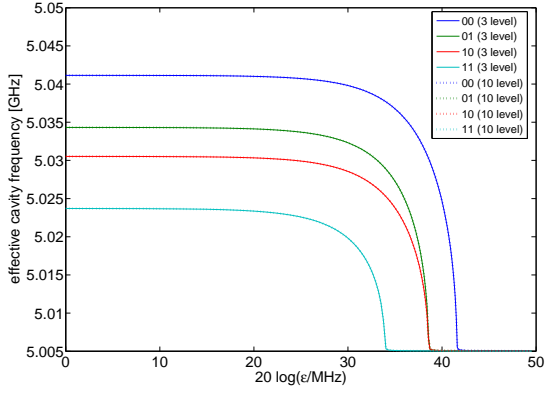}
\caption{\label{fig:higher_levels} Comparison between the results of the effective cavity frequency when including three or ten energy levels. We see that the results are absolutely identical, such that it seems to be reasonable to only include the lowest non occupied energy level into the calculations to get correct results. The parameters are the same as in Fig. \ref{fig:2QB_1}.}
\end{figure}

In this section of the appendix we want to show that only the lowest non occupied energy level (in our case $\ket{2}$) of the qubits has an effect on the results. Including even higher levels does not change either the frequency shifts nor the behavior of the photon amplitude. This can be seen in Fig. \ref{fig:higher_levels} where we show the effective frequency shift and compare the results when we include the lowest three energy levels with the results when we include the lowest ten. One sees that the results of the two cases are completely identical, such that we can claim that only the three lowest levels affect the results. More precisely it seems reasonable that the lowest non occupied energy level is the last one that has an effect on the system.

\section{Proof for exact dispersive transformation}
\label{app:5}
In this section of the appendix we will prove two relations for nested commutators that we need to split up terms when we perform the exact dispersive transformation including more than two energy levels.

\subsection{First proof}

Given four operators $A$, $B$, $C$, and $D$, we want to show that
\begin{align}
\textbf{ad}_{A+B}^n \left(B+C\right) = \textbf{ad}_A^n(C) + \textbf{ad}_B^n(D),
\label{proof1}
\end{align} 
if the following relation is fulfilled:
\begin{align}
[A,B] = [A,D] = [B,C] = [C,D] = 0.
\label{rel1}
\end{align}
We use induction to show that \eqref{proof1} is valid for all $n$. Let's start with the case $n=1$ which is trivial:
\begin{align}
[A+B,C+D] \underbrace{=}_{\eqref{rel1}} [A,C] + [B,D].
\label{induction1} 
\end{align}
So we now that for $n=1$ the relation holds. Lets assume that for $n=k$ \eqref{induction1} holds, that is 
\begin{align}
\textbf{ad}_{A+B}^k(C+D) = \textbf{ad}_A^k(C) + \textbf{ad}_B^k(D).
\end{align}
Let $n=k+1$:
\begin{align}
\begin{split}
\textbf{ad}_{A+B}^{k+1}(C+D) &= \left[A+B,\textbf{ad}_{A+B}^k(C+D)\right] \\
&= \left[A+B,\textbf{ad}_A^k(C)+\textbf{ad}_B^k(D)\right] \\
&= \left[A,\textbf{ad}_A^k(C)\right] + \left[B,\textbf{ad}_B^k(D)\right] \\&+ \left[A,\textbf{ad}_B^k(D)\right] + \left[B,\textbf{ad}_A^k(C)\right].
\end{split}
\end{align}
To get the relation we want, we have to show that
\begin{align}
\left[A,\textbf{ad}_B^k(D)\right]=\left[B,\textbf{ad}_A^k(C)\right]=0.
\end{align}
For this we again use induction and only show it for one of the terms, since the second calculation is analog. Again the case $n=1$ is fulfilled:
\begin{align}
\left[A,[B,D]\right] = - \left[B,[D,A]\right] - \left[D,[A,B]\right] \underbrace{=}_{\eqref{rel1}} 0, 
\end{align}
where we used the Jacobi-identity for operators. So lets assume the statement is true for $n=k$. Let $n=k+1$
\begin{align}
&\left[A,\textbf{ad}_B^{k+1}(D)\right] = \left[A,\left[B,\textbf{ad}_B^k(D)\right]\right] \\
&\hspace{0.2cm}= - \left[B,\left[\textbf{ad}_B^k(D),A\right]\right] - \left[\textbf{ad}_B^k(D),\left[A,B\right]\right]=0
\end{align}
where we again used the Jacobi-identity and the induction hypothesis. In the same manner we can show that $\left[B,\textbf{ad}_A^k(C)\right]=0$, such that we finally proofed \eqref{induction1} under the condition \eqref{rel1} for all $n\in \mathbbm{N}$.
\subsection{Second Proof}
Here want to prove a second identity we need to perform the exact dispersive transformation in our case. We show that
\begin{align}
\textbf{ad}_{A+B}^k(C) = \textbf{ad}_A^k(C) + \textbf{ad}_B^k(C),
\label{induction2}
\end{align}
if the following relations are satisfied:
\begin{align}
[A,B] &= 0 \label{rel2}\\
\left[B,[C,A]\right] &= \left[A,[C,B]\right] = 0.
\label{rel3}
\end{align} 
Again we use induction. The identity is trivial to show for $n=1$. So we assume \eqref{induction2} holds for $n=k$. Let $n=k+1$:
\begin{align}
\textbf{ad}_{A+B}^{k+1}(C) &= \left[A+B,\textbf{ad}_{A+B}^k(C)\right] \\
&= \left[A+B,\textbf{ad}_A^k(C) + \textbf{ad}_B^k(C)\right] \\
&= \textbf{ad}_A^{k+1}(C) + \textbf{ad}_B^{k+1}(C)\\&\hspace{0.3cm}+ \left[A,\textbf{ad}_B^k(C)\right] + \left[B,\textbf{ad}_A^k(C)\right].
\end{align}
To prove \eqref{induction2} we therefore have to show that 
\begin{align}
\left[A,\textbf{ad}_B^k(C)\right] =  \left[B,\textbf{ad}_A^k(C)\right] = 0. 
\end{align}
We start with the first term. For $n=1$:
\begin{align}
\left[A,[B,C]\right] = - \left[B,[C,A]\right] - \left[C,[A,B]\right] = 0,
\end{align}
where we used the Jacobi-identity and relations $\eqref{rel2}$ and \eqref{rel3}. Let's assume we have proven the identity for $n=k$. Let $n=k+1$:
\begin{align}
\left[A,\textbf{ad}_B^{k+1}(C)\right] &= \left[A,\left[B,\textbf{ad}_B^k(C)\right]\right] \\ 
= -\left[B,\left[\textbf{ad}_B^k(C),A\right]\right]&-\left[\textbf{ad}_B^k(C),\left[A,B\right]\right] = 0
\end{align}
Likewise one can show that $\left[B,\textbf{ad}_A^k(C)\right]=0$ and therefore we have proven the identity \eqref{induction2} under the conditions \eqref{rel2} and \eqref{rel3} for all $n\in \mathbbm{N}$.

\section{Equation system to determine dephasing}
Here we solve the equation system to get the expression for the cavity leakage induced dephasing of Sec. \ref{sec:4_1}. Putting the density matrix \eqref{deph:density} into the Lindblad equation \eqref{deph:Lindblad} we get equations of motion for the cavity parts of the density matrix
\begin{widetext}
\begin{align}
\dot{\hat\rho}_{11} &= \kappa \mathcal{D}[\hat a]\hat\rho_{11} - (\gamma_1^{(1)}+\gamma_1^{(2)})\hat\rho_{11} -i\epsilon\left[\hat a^{\dag}+\hat a,\hat\rho_{11}\right]-i(\chi_{1}^{(1)}+\chi_1^{(2)})\left[\hat a^{\dag}\hat a,\hat \rho_{11}\right]-i\delta_c\left[\hat a^{\dag}\hat a,\hat\rho_{11}\right]\label{den_system1} \\
\dot{\hat\rho}_{00} &= \kappa \mathcal{D}[\hat a]\hat\rho_{00} + (\gamma_1^{(1)}+\gamma_1^{(2)})\hat\rho_{11} -i\epsilon\left[\hat a^{\dag}+\hat a,\hat\rho_{00}\right]+i(\chi_1^{(1)}+\chi_1^{(2)})\left[\hat a^{\dag}\hat a,\hat \rho_{00}\right]-i\delta_c\left[\hat a^{\dag}\hat a,\hat\rho_{00}\right]\label{den_system2} \\
\begin{split}
\dot{\hat\rho}_{10} &= \kappa \mathcal{D}[\hat a]\hat\rho_{10} - (\gamma_2^{(1)}+\gamma_2^{(2)})\hat\rho_{10} -i\epsilon\left[\hat a^{\dag}+\hat a,\hat\rho_{10}\right]\\&-i\left[(\chi_1^{(1)}+\chi_1^{(2)})\left\{\hat a^{\dag}\hat a,\hat\rho_{10}\right\}-(\chi_2^{(1)}+\chi_2^{(2)})\hat a^{\dag}\hat a \hat\rho_{10}\right]-i\delta_c\left[\hat a^{\dag}\hat a,\hat\rho_{10}\right] -i(\tilde\omega_1+\frac{\tilde\omega_2}{2})\hat \rho_{10} \end{split}\label{den_system3}\\
\begin{split}
\dot{\hat\rho}_{01} &= \kappa \mathcal{D}[\hat a]\hat\rho_{01} - (\gamma_2^{(1)}+\gamma_2^{(2)})\hat\rho_{01} -i\epsilon\left[\hat a^{\dag}+\hat a,\hat\rho_{01}\right]\\&+i\left[(\chi_1^{(1)}+\chi_1^{(2)})\left\{\hat a^{\dag}\hat a,\hat\rho_{10}\right\}- (\chi_2^{(1)}+\chi_2^{(2)})\hat\rho_{10}\hat a^{\dag}\hat a\right] - i\delta_c\left[\hat a^{\dag}\hat a,\hat\rho_{01}\right]+i(\tilde\omega_1+\frac{\tilde\omega_2}{2})\hat \rho_{01},
\end{split}\label{den_system4}
\end{align}
\end{widetext}
where $\gamma_2^{(j)}=\gamma_{1}^{(j)}+\gamma_{\Phi}^{(j)}/2$ and $\chi_i^{(j)} = (g_i^{(j)})^2/\tilde \Delta_i^{(j)}$ . Now we assume that both qubits have the same relaxation and dephasing rate and define $\tilde\omega = \tilde\omega_1+\tilde\omega_2/2$. In general there exist no solution for these four equations because of the coupling term introduced by $\gamma_1$. In our case we are only interested in dephasing rate, such that we can set $\gamma_1=0$ in the equations for the diagonal parts. To solve the above equation system, we consult the generalized P representation and express the cavity density matrix elements as
\begin{align}
\hat \rho_{ij}^c = \int \Lambda(\alpha,\beta)P(\alpha,\beta){\rm d}\mu(\alpha,\beta),
\end{align} 
with probability densities $P_{ij}$. Here we use the so called positive-P representation, where
\begin{align}
\Lambda(\alpha,\beta)&= \frac{\ket{\alpha}\bra{\beta^*}}{\braket{\beta^*|\alpha}} \\
{\rm d}\mu(\alpha,\beta)&= {\rm d}^2\alpha{\rm d}^2\beta
\end{align}
Putting the positive-P representation of the matrix elements into equations \eqref{den_system1}-\eqref{den_system4} using the relations (see e.g. \cite{GardinerZoller})
\begin{align}
\hat a\Lambda(\alpha,\beta) &= \alpha\Lambda(\alpha,\beta)\\
\hat a^{\dag}\Lambda(\alpha,\beta) &= (\beta+\partial_{\alpha})\Lambda(\alpha,\beta)\\
\Lambda(\alpha,\beta)\hat a^{\dag} &= \beta \Lambda(\alpha,\beta)\\
\Lambda(\alpha,\beta)\hat a &= (\partial_{\beta} + \alpha)\Lambda(\alpha,\beta)
\end{align}
we get equations of motion for these probability densities
\begin{widetext}
\begin{align}
\dot P_{11} &= \partial_{\alpha} \left[(i\epsilon + 2i(\chi_{1}-\chi_2)\alpha + i\delta_c\alpha +\kappa\alpha/2)P_{11}\right]+\partial_{\beta}\left[(-i\epsilon-2i(\chi_{1}-\chi_2)\beta -i\delta_c\beta+\kappa\beta/2)P_{11}\right] \\
\dot P_{00} &= \partial_{\alpha} \left[(i\epsilon - 2i\chi_{1}\alpha + i\delta_c\alpha +\kappa\alpha/2)P_{00}\right]+\partial_{\beta}\left[(-i\epsilon+2i\chi_{1}\beta -i\delta_c\beta+\kappa\beta/2)P_{00}\right] \\
\begin{split}
\dot P_{10} &= \partial_{\alpha} \left[(i\epsilon + 2i(\chi_1-\chi_2)\alpha + i\delta_c\alpha +\kappa\alpha/2)P_{10}\right]+\partial_{\beta}\left[(-i\epsilon+2i\chi_1\beta -i\delta_c\beta+\kappa\beta/2)P_{10}\right] \\
&\hspace{0.2cm}-i4(\chi_1-\chi_2/2) \alpha\beta P_{10} - 2\gamma_2 P_{10} - i\tilde\omega P_{10} \end{split}\\
\begin{split}
\dot P_{01} &= \partial_{\alpha} \left[(i\epsilon - 2i\chi_1\alpha + i\delta_c\alpha +\kappa\alpha/2)P_{01}\right]+\partial_{\beta}\left[(-i\epsilon-2i(\chi_1-\chi_2)\beta -i\delta_c\beta+\kappa\beta/2)P_{01}\right] \\
&\hspace{0.2cm}+i4(\chi_1-\chi_2/2) \alpha\beta P_{01} - 2\gamma_2 P_{10} + i\tilde\omega P_{01} 
\end{split}
\end{align}
\end{widetext}
Here we assumed identical qbuits, hence $\chi_i^{(j)}=\chi_i^{(k)}=\chi_i$ and $\gamma_i^{(j)}=\gamma_i^{(k)}=\gamma_i$. These equations can be solved with the Ansatz
\begin{align}
P_{11} &= \delta^{(2)}\left[\alpha-\alpha_1(t)\right]\delta^{(2)}\left[\beta-\alpha_1^*(t)\right] \\
P_{00} &= \delta^{(2)}\left[\alpha-\alpha_0(t)\right]\delta^{(2)}\left[\beta-\alpha_0^*(t)\right]\\
P_{10} &= a_{10}(t)\delta^{(2)}\left[\alpha-\alpha_1(t)\right]\delta^{(2)}\left[\beta-\alpha_0^*(t)\right]\\
P_{01} &= a_{01}(t)\delta^{(2)}\left[\alpha-\alpha_0(t)\right]\delta^{(2)}\left[\beta-\alpha_1^*(t)\right]
\end{align}
which yields the following differential equations:
\begin{align}
\dot\alpha_1 &= -i\epsilon -i\left(\delta_c + \chi_1-\chi_2 -i\kappa/2\right)\alpha_1 \label{dg1}\\
\dot \alpha_0 &= -i\epsilon -i\left(\delta_c -\chi_1 -i\kappa/2\right)\alpha_0\label{dg2}\\
\dot a_{10} &= -i(\tilde \omega -i2\gamma_2)a_{10} - i4(\chi_1-\chi_2/2)\alpha_1\alpha_0^*a_{10}\label{dg3}\\
\dot a_{01} &= i(\tilde \omega +i2\gamma_2)a_{01} +i4(\chi_1-\chi_2/2)\alpha_0\alpha_1^*a_{01}\label{dg4}.
\end{align}
In this equation system we see the phase difference with which the two states $\ket{\alpha_0}$ and $\alpha_1$ oscillate, which leads to an effective dephasing. The differential equations for $\alpha_i$ and $a_{ij}$ can easily be solved and lead to the time evolution of the density matrix \eqref{deph:density_time} we used in Sec \ref{sec:4_2} to calculate the respective dephasing rate
\end{appendix}

\bibliography{Bibliography.bib}
\bibliographystyle{apsrev4-1}

\end{document}